\documentclass[twocolumn,preprintnumbers,showpacs,amsmath,amssymb]{revtex4}

\usepackage{psfrag}
\usepackage{graphicx}
\usepackage{dcolumn}
\usepackage{bm}
\usepackage{epsfig}
\usepackage[notcite,notref]{}

\begin{document}

\title{A quantum Boltzmann equation for intrinsic decoherence in a dilute gas}

\author{L. Rico-P\'erez}
\author{J.R. Anglin}
\affiliation{Physics Department and State Research Center OPTIMAS, University of Kaiserslautern,
Erwin-Schr\"odinger-Stra\ss e. 46, D-67663 Kaiserslautern, Germany}

\date{\today}

\begin{abstract}
We use a careful treatment of time-dependent wave-mechanical scattering to determine the conditions under which a dilute, non-degenerate quantum gas can obey a Boltzmann equation. If the gas possesses weak long-range coherence, such as may occur when a gas is quantum mechanically throttled by tunnelling through a barrier, collisions within the gas will suppress this long range coherence. We calculate its decay rate and find it to be related to the Boltzmann equation's loss term.
\end{abstract}


\maketitle

\section{Introduction}

The Boltzmann equation is a semi-phenomenological description of non-equilibrium gas dynamics. It is one of the most useful equations in physics, but it is also one of the most embarrassing, because it has neither been established as a first principle in its own right nor derived rigorously from first principles. Instead it has stood for nearly 150 years as a paradigm of intuition in physics, while the first principles that were accepted in Boltzmann's time have all been replaced. The fundamental description of gas dynamics is now quantum many-body theory, but of the many different equations that are now used to describe different kinds of non-equilibrium gas dynamics, most can still be recognized as quantum generalizations of the original Boltzmann equation.

This is not mere conceptual inertia, because the basic idea of a Boltzmann-like equation assumes a coarse-grained description of a many-body system, even if it is far from equilibrium. Since the coarse-grained description aims to include only those properties of a gas which are macroscopically measurable, there is a built-in tendency to think in terms of classical quantities, even if quantum effects are allowed to alter the time evolution of those quantities. In principle, however, quantum gases can evolve out of equilibrium in radically non-classical ways, such as by tunnelling. A macroscopic volume of gas could in principle exist in a state with long-range quantum coherence, and the textbook explanations of why such quantum effects are not seen macroscopically would not obviously apply when the individual particles in the gas are all microscopic. In this paper we show how a Boltzmannian approach which allows such non-classical gas behavior leads to a form of master equation describing intrinsic quantum decoherence in a gas due to collisions among the particles which compose the gas itself, thus showing how interactions among many gas particles can suppress collective quantum behavior. Our results will also make clear how this decoherence effect relates to the classically familiar kinds of non-equilibrium evolution---relaxation and thermalization---that are usually described in Boltzmannian terms. 

We will start from the observation that Boltzmann-like equations are based on two heuristic approximations. The approximation that is most widely recognized---the neglect of correlations that is known as the \textit{Stosszahlansatz}---actually depends on a prior approximation that represents the interaction between two particles as an instantaneous event at a point. While this approximation of instantaneous collisions is in many cases excellent, even quantum mechanically, making it too naively can mean overlooking important quantum phenomena right from the start.

Our derivation will reverse the order of these two Boltzmannian approximations. We will pass quickly and heuristically from the many-body problem to the two-body problem, just by appealing naively to diluteness. We will then look more carefully at the range of length and time scales that are involved in quantum two-body scattering, and show how a consistently quantum mechanical approximation based on this scale hierarchy leads to a Boltzmann equation for the more classical features of the quantum gas's single particle distribution. For the long-range spatial coherence of the gas, however, this same consistent approach leads instead to a master equation describing collisional decoherence. Our paper is somewhat long, and much of it consists of pedagogical review of long-established results, but we ask the interested reader to have patience. Some of the known results that we review are just a short step beyond standard textbook treatments, and hence although they should certainly be known, they are not widely known. Quantum many-body dynamics is more complicated that it can be made to seem by rushing through it, and fundamentally significant questions can be overlooked when one leaps too quickly to an anticipated conclusion.

The paper is organized as follows. We begin immediately below by briefly reviewing the classical Boltzmann equation, as well as some of its quantum generalizations. We also explain the role of time scale hierarchy in Boltzmann's theory. In Section III we then introduce a formally exact description of gas dynamics using a second-quantized Heisenberg picture, and reduce it to an actually solvable two-body problem by postulating a quantum version of the \textit{Stosszahlansatz}. In Section IV we will apply our two-body theory in two situations: the first one to confirm that it leads, with certain additional assumptions and approximations, to the Boltzmann equation for the single-particle Wigner function of the gas; and the second one to show that the two-body theory from Section III also implies collisional decoherence, which is our main result. We conclude in Section V with a brief discussion and outlook toward future work.

\section{Boltzmann Theory}
\subsection{The original Boltzmann equation}
Statistical mechanics is applied to systems composed of too many particles to be described precisely. Only a {\it coarse-grained} representation of such a system is feasible or even desirable. In classical statistical mechanics the coarse-grained description is usually based on the single-particle phase space distribution function $f(\mathbf{r},\mathbf{p},t)$, representing the marginal probability density of finding any one particle at phase space point $(\mathbf{r},\mathbf{p})$, regardless of where all the other particles are and what momenta they carry. In an external potential $V(\mathbf{r})$ and in the absence of interactions between the particles, the distribution function $f$ evolves in time according to Liouville's equation:
\begin{eqnarray}\label{eq:Liouville}
\partial_t f(\mathbf{r},\mathbf{p},t) &=&  - \frac{\mathbf{p}}{m}\cdot\nabla_r f + \nabla_r V \cdot \nabla_p f\;.
\end{eqnarray}

The classical Boltzmann equation (CBE) incorporates a particular class of effects due to short-ranged interactions between particles, by adding a collision term
$$\Gamma[f](\mathbf{r},\mathbf{p}) =  \Gamma_{+}-\Gamma_{-}$$
to the right side of the Liouville equation (\ref{eq:Liouville}). $\Gamma_{+}$ represents the effect of a particle colliding with another particle at position $\mathbf{r}$, such that the two particles emerge with momenta $\mathbf{p}$ and any other arbitrary momentum $\mathbf{p}^{\prime}$, respectively; while $\Gamma_{-}$ describes the deflection of a particle at $\mathbf{r}$ away from initial momentum $\mathbf{p}$, after hitting another particle with arbitrary initial momentum $\mathbf{p}^{\prime}$. In both cases it is taken for granted that particles barely move from the initial position $\mathbf{r}$ during the collision, so that the $\Gamma_\pm$ terms represent collisional scattering as a redistribution of momentum between two particles which is completed in a single instant of time and at a single point in space. Both total momentum and total kinetic energy are conserved in the collision process, and so if the collision has momentum transfer $\mathbf{p}^{\, \prime\prime}$, then the collision terms are explicitly
\begin{widetext}
\begin{eqnarray}\label{eq:CSAGain}
\Gamma_+ &=& \frac{1}{m}\int\!d^3p'd^3p''\, \delta\Bigl(|\mathbf{p}^{\, \prime\prime}|^2-\frac{|\mathbf{p}-\mathbf{p}^{\prime}|^2}{4}\Bigr)\,\frac{d\sigma}{d\Omega}\, f\left(\mathbf{r},\frac{\mathbf{p}+\mathbf{p}^{\prime}}{2}+\mathbf{p}^{\, \prime\prime},t\right)\, f\left(\mathbf{r},\frac{\mathbf{p}+\mathbf{p}^{\prime}}{2}-\mathbf{p}^{\, \prime\prime},t\right)
\end{eqnarray}
\begin{eqnarray}\label{eq:CSALoss}
\Gamma_- &=& \frac{1}{m}\int\!d^3p'd^3p''\, \delta\Bigl(|\mathbf{p}^{\, \prime\prime}|^2-\frac{|\mathbf{p}-\mathbf{p}^{\prime}|^2}{4}\Bigr)\,\frac{d\sigma}{d\Omega}\, f(\mathbf{r},\mathbf{p},t)\, f(\mathbf{r},\mathbf{p}^{\,\prime},t)\;.
\end{eqnarray}
\end{widetext}
The differential cross section traditionally denoted by $\frac{d\sigma}{d\Omega}$ is simply a function of momentum transfer $\mathbf{p}^{\, \prime\prime}$ which expresses the rate at which collisions with the given momentum transfer occur, and the total cross section is simply its integral over all possible directions  $\sigma (q) = \int d\Omega\,\frac{d\sigma(\mathbf{q})}{d\Omega} $.

Alternatively we can express the loss term (\ref{eq:CSALoss}) in a more compact way as 
\begin{eqnarray}\label{eq:CSALossCompact}
\Gamma_- &=& \gamma(\mathbf{r},\mathbf{p},t)\, f(\mathbf{r},\mathbf{p},t) \\
\gamma(\mathbf{r},\mathbf{p},t) &=& \int\!d^3q\,\frac{q}{m} \frac{\sigma(q)}{4\pi}\,  f(\mathbf{r},\mathbf{p}+2\mathbf{q},t)\;.
\end{eqnarray}
This form emphasizes that $\gamma$ effectively plays the role of a probability decay rate $\partial_t f = -\, \gamma\, f\, +\,...$, and can be considered to set the time scale for collisional equilibration.

The CBE may at first seem intuitive to the point of being obvious, but it becomes less obvious if one stops to think more about it. According to Eqns.~(\ref{eq:CSAGain}) and (\ref{eq:CSALoss}), collisions change the phase space probability distribution $f$ continuously in time, and locally in position; but in general the CBE provides change in momentum $\mathbf{p}$ that is \textit{non-local} in phase space. For example, suppose a distribution $f$ initially has support only for two opposite momenta $\pm \mathbf{p}_{0}$
\begin{eqnarray}\label{eq:DummyInitialOppositeMomenta}
f(\mathbf{x},\mathbf{p},t=0) \propto \sum_\pm \delta^3(\mathbf{p}\mp \mathbf{p}_0)
\end{eqnarray}
The $\Gamma_{+}$ term in the CBE will immediately generate support for $f$ over the entire momentum-sphere $|\mathbf{p}|=|\mathbf{p}_{0}|$
\begin{eqnarray}\label{eq:DummyScatteredOppositeMomenta}
f(\mathbf{x},\mathbf{p},\Delta t) \approx  f(\mathbf{x},\mathbf{p},0)  + A(\mathbf{x},\mathbf{p}) \,\delta(p-p_0)\, \Delta t .
\end{eqnarray}
This support will not grow by spreading out from the two initial probability spikes at $\pm\mathbf{p}_{0}$, however; it will fade into place over the entire sphere at once, weighted by the angular dependence of $\frac{d\sigma}{d\Omega}$. This distinctly non-Liouvillian behavior of $f$ under the CBE occurs because the CBE represents collisions as instantaneous changes of momentum by finite amounts, and so probability can grow even at momenta which have no probability near them. In reality, interacting particles change their momenta continuously in response to finite forces, but the Boltzmannian description of collisions idealizes continuous momentum change under strong inter-particle forces as instantaneous jumps in momentum.

Although this radically non-Liouvillian idealization of collisions is essential to Boltzmannian theory, the CBE is more notorious for a different idealization: the \textit{Stosszahlansatz} (``collision number hypothesis"), which places the simple product $f(\mathbf{r},\mathbf{p},t) f(\mathbf{r},\mathbf{p}',t)$ in the integrand of $\Gamma_{-}$, as well as the similar product of two $f$ functions in $\Gamma_{+}$. What should more rigorously appear in these integrals are not these products of two single-particle distributions, but rather a single two-particle distribution, $f_{2}(\mathbf{r},\mathbf{p};\mathbf{r},\mathbf{p}';t)$. The heuristic assumption that this $f_{2}$ can be thus factorized is the classical \textit{Stosszahlansatz}. 

The \textit{Stosszahlansatz} assumption that two-particle correlations can be ignored is discussed much more often than the instantaneous collision model, but in fact the \textit{Stosszahlansatz} depends on instantaneous collisions in an essential way. To ignore correlations during a collision is absurd, because a collision \textit{is} nothing but the fact that two nearby particles influence each other's motion. Hence it is only possible to discuss collisional effects without mentioning correlations if one represents the entire collision duration as an infinitesimal instant during which one uncorrelated state of the system is instantaneously replaced by a another state that is also uncorrelated. This macroscopic picture of collision effects can be accurate within its own terms, but only if the instantaneous collision model correctly represents the long-term, large-scale effects of the actual two-body scattering. Ensuring that this is so is not trivial, even in classical mechanics; quantum mechanical scattering introduces further complications as well.

\subsection{Quantum Boltzmann equations}
As a partial integro-differential equation for a function in seven dimensions (time plus six-dimensional phase space), the CBE is notoriously difficult to solve; but it is still enormously easier to solve than the full quantum many-body problem, which is posed in a Hilbert space of enormous dimension. Because many-body Hilbert spaces are so large, however, there are many possible ways of simplifying the many-body problem into some form of quantum Boltzmann equation (QBE). Several quite different kinds of QBE have therefore been proposed. 

One basic version of QBE modifies the CBE only by inserting additional factors of $f(\mathbf{r},\mathbf{p},t)$ in the scattering terms, to represent Pauli exclusion or bosonic enhancement of scattering into occupied states, on the assumption that the CBE is otherwise valid as it is \cite{Nord28} even though the classical phase space picture is fundamentally inconsistent with Heisenberg uncertainty. One may also add a mean-field term, producing what has been called a Vlasov equation for neutral atoms \cite{Cercignani75}
\begin{eqnarray}\label{BVE}
\partial_tf(\mathbf{r},\mathbf{p},t) &=&  - \frac{\mathbf{p}}{m}\cdot\nabla_r f +  \Gamma_{+}[f]-\Gamma_{-}[f]\nonumber\\
&&+ \mathbf{\nabla}_{r}V_{\rm mf}\cdot\mathbf{\nabla}_{p}f\;,
\end{eqnarray}
where the mean-field potential $V_{\mathrm{mf}}(\mathbf{r},t)$ is proportional to the gas density $\int\!d^{3}p\,f$ at $\mathbf{r}$, and to the real part of the scattering amplitude for collisions (in the limit of cold collisions, to the s-wave scattering length). Although the mean-field potential is determined by the gas itself, it has the same dynamical effect as an external conservative potential applied to the gas particles. It is a Hamiltonian term which does not provide dissipation or equilibration.

A more thoroughly quantum mechanical revision of the Boltzmann equation \cite{Tolm38} replaces the basic $f(\mathbf{q},\mathbf{p},t)$ with the expected occupation numbers of single-particle energy eigenstates $f(E,t) = \mathrm{Tr}[\hat{\rho}(t) \hat{n}_E]$, where $\hat{\rho}(t)$ is the density matrix in the many-body Hilbert space, describing a time-dependent mixed quantum state of the entire many-particle system. The assumption that the restricted set of expectation values $f(E,t)$ can adequately describe even just the single-particle properties of the system is a restriction to a special case, however. The probability distribution $f(E,t)$ does not contain information about coherences between different energy eigenstates; it describes only incoherent mixtures of the global energy eigenstates, and hence it can only describe departures from equilibrium that are similarly global. If the energy eigenstates are plane waves, for example, then $f(E,t)$ describes only states that may be out of equilibrium in the sense that they are not canonical ensembles, but that are still homogeneous in the sense that they are invariant under spatial translation. This is in contrast to the way the CBE can describe non-equilibrium $f$ that are arbitrary functions of $\mathbf{r}$ as well as $\mathbf{p}$. The severity of this restriction to global states is often glossed over in papers that purport to derive `the' quantum Boltzmann equation \cite{Greenwood}. 

To go beyond $f(E,t)$ in one direction, sophisticated quantum kinetic theories have been derived (see for example \cite{Gard97, ZNGr98}) to describe time-dependent number fluctuations and coherence among multiply-occupied single-particle states in degenerate gases. This class of theories involves separate time-dependent probabilities for each possible occupation number of each bosonic mode, $p(\{n_{j}\},t)$, and not just an average occupation number $f(E,t)$. To simplify this complex problem, approximations based on some partial form of local equilibrium have generally been made. In particular, it has often been assumed that the non-equilibrium process in question is the time-dependent growth of a Bose-Einstein condensate under evaporative cooling, in a (so-called) collisionless regime. For such a process, the detailed distribution of occupation numbers in low-energy states is very important, but spatial locality and spatial coherence among non-condensate modes is typically not.

In this paper we go beyond previous quantum Boltzmann equations in a different direction. We will restrict our attention to quantum gases that are far from quantum degeneracy; in other words, we will assume that all single-particle quantum states have only small probabilities of being occupied at all, and no chance of being multiply occupied. There is more to the quantum nature of quantum gases than quantum degeneracy, however. Even if the density of gas particles is well below one per cubic thermal de Broglie wavelength, each particle is still a quantum mechanical particle, capable of non-classical behavior such as interference or tunnelling. We will consider quantum gases that may be far from equilibrium in such dramatically non-classical ways. We will recover the classical Boltzmann equation for gases in quasi-classical states, but for gases in severely non-classical states, such as may be produced when a dilute gas tunnels through a potential barrier, we will obtain a non-Boltzmannian master equation which describes an essentially quantum mechanical aspect of dilute gas kinetics: intrinsic decoherence due to collisions.

\subsection{Reconsidering collisions}
We will not obtain this insight by reconsidering the \textit{Stosszahlansatz}. Instead we will make the standard kind of factorization assumptions, with the same heuristic justification (or lack thereof). Our focus instead will be on the more basic Boltzmannian idealization, whereby collisions were represented as instantaneous redistributions of momentum. We will reconsider quantum mechanical two-body collisions as evolutions over distances and time intervals which are long compared to interaction length and time scales, but short compared to the length and time scales over which the macroscopic features of the gas can significantly change. 

The model of an instantaneous collision is especially subtle for quantum mechanical collisions, because the differential cross section which characterizes collisional scattering is formally defined as a relationship between asymptotic pre- and post-interaction states separated by \textit{infinite} time. Textbook scattering theory is explained in terms of energy eigenstates, which are time-independent steady states; so far from being instantaneous events, collisions would thereby seem to be eternal.
\begin{figure}
\begin{center}
\includegraphics[width=0.5\textwidth]{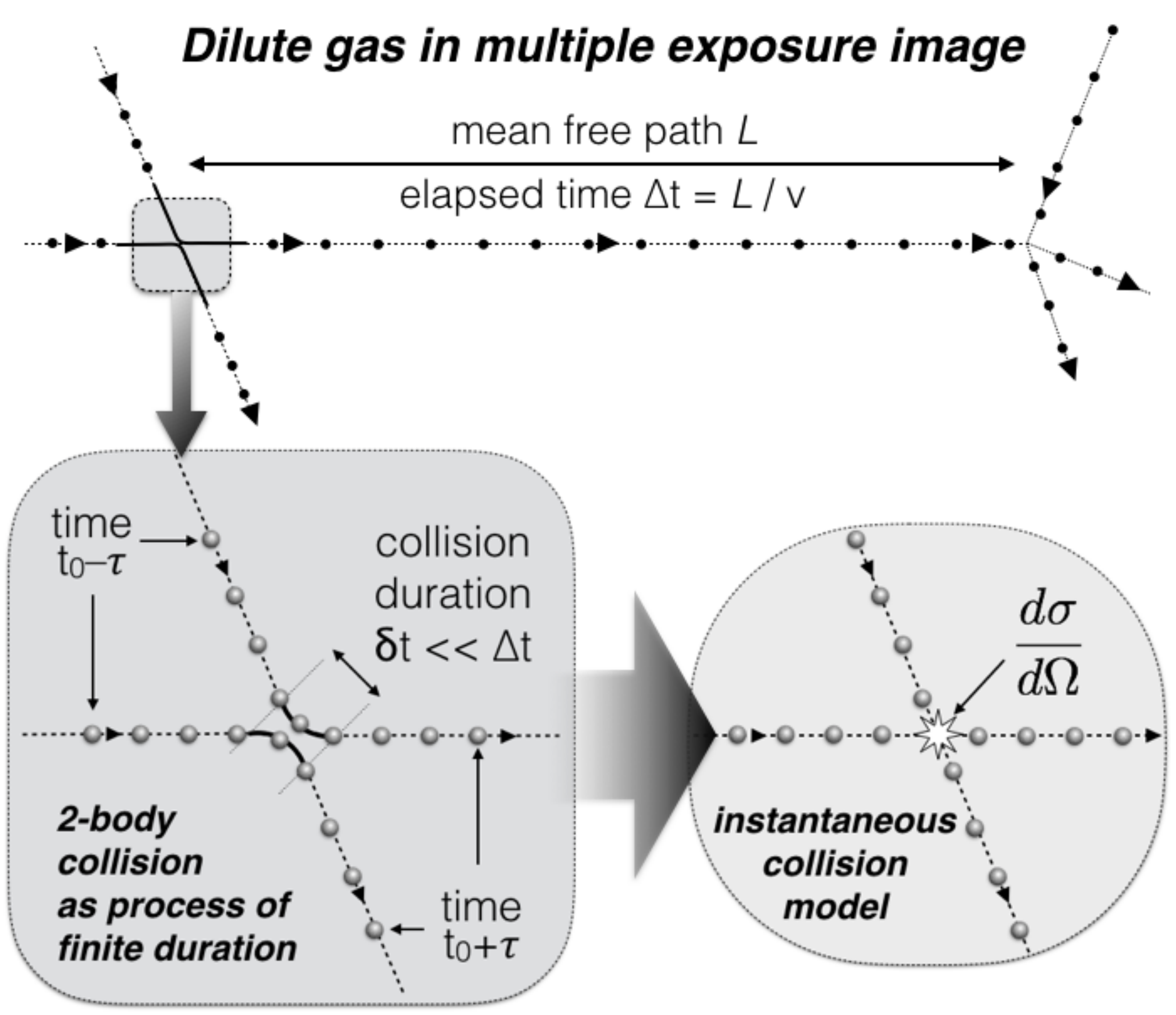}
\end{center}
\caption{Schematic representations of gas particles moving and colliding, as if in a multiple exposure photograph. If the durations $\delta t$ over which particles interact are much shorter than the typical time between collisions $\Delta t$ then there exists an intermediate time scale $\tau = \sqrt{\delta t\,\Delta t}$ which is simultaneously short compared to $\Delta t$ and long compared to $\delta t$. For gas evolution on time scales of order $\Delta t$, a collision that takes place within a time interval of order $\tau$ can be modeled as \textit{instantaneous} ($\tau/\Delta t\to 0$) event at time $t_{0}$, even thought it is characterized by a differential cross section $d\sigma/d\Omega$ which describes \textit{asymptotic} two-body evolution ($\tau/\delta t\to\infty$).}
\label{fig:scales}
\end{figure}

The resolution of this apparent paradox is shown in Fig.~1. The time interval within which two colliding particles in a gas are effectively interacting will have some typical \textit{duration} time scale $\delta t$.  The typical time between the collision of any given particle with some second particle, and the next collision of the first particle with some third particle--- \textit{i.e.}, the intercollision time---will be some generally different time scale $\Delta t$. In a dilute gas, the ratio between these two time scales will be some small $\delta t/\Delta t = \epsilon \ll 1$. We will therefore be able to define an intermediate time scale $\tau \sim \sqrt{\delta t\,\Delta t}$. With this constructed intermediate time scale $\tau$ as a sort of auxiliary variable, we will then be able to describe the collision itself as two-body evolution over a time interval $\tau = \epsilon^{-1/2}\delta t = \sqrt{\epsilon}\Delta t$. When we calculate to leading order in $\epsilon$, the duration $\tau$ of two-body interaction is indeed simultaneously infinite and infinitesimal, because for $\epsilon\to 0$ we have $\tau/\delta t\to \infty$ and yet $\tau/\Delta t \to 0$. This time scale hierarchy $\delta t \ll \tau \ll \Delta t$ is a defining feature of dilute gases. A system in which interactions between two particles do not happen fast enough to ignore the intervention of further particles will fall into the categories of dense gases or liquids, which are much more complicated.

A dilute gas's various time scales all have associated length scales, which will also be important in the gas dynamics. The intercollision time $\Delta t$ can be associated with the {\it mean free path} $L$; this distance must be much larger than the {\it interaction range}, which we could associate with the collision duration $\delta t$; and finally, we can denominate the distance covered by a particle that flies during our intermediate time scale $\tau$ around a collision as the {\it scattering reach}. 

It should be noted that all of these length and time scales represent characteristic orders of magnitude and not precise quantities. The intermediate time scale $\tau$ and the associated scattering reach in particular may be arbitrary to within orders of magnitude. In air at room temperature, for example, the typical particle speed is on the order of 300 m/s while the interaction ranges are fractions of nanometers; so we can estimate $\delta t$ around $10^{-12}$ s. The typical particle density is of order $10^{25}$ m$^{-3}$, so the mean free path is of order $L\sim 10^{-6}$m and the intercollision time $\Delta t\sim 10^{-8}$s. For air at room temperature therefore our intermediate time $\tau$ could thus be anywhere around $10^{-10}$s, and our scattering reach on the order of $10^{-8}$m. The precise value of $\tau$ is not meaningful but the time scale hierarchy $\delta t\ll\tau\ll\Delta t$ is an important quantitative fact. In later Sections we will see that certain gas distribution functions must vary only slowly and slightly over the scattering reach distance scale that is associated with the intermediate time scale $\tau$, or else collisions become inherently non-local events that cannot be described in simple Boltzmannian terms. This is a somewhat stronger requirement than simply stipulating that the gas properties vary slowly on the scale of the interaction range: the scale of their spatial variation must be so much longer than the interaction range that we can insert a whole new length scale between them, compared to which the variation scale is large enough to be approximated in leading order as infinite, while the interaction range can be approximated as infinitesimal.

\section{Classical Boltzmann equation for a dilute quantum gas: derivation from explicit scattering dynamics}\label{Section2ndQ}

In this Section we will review how the neglect of three-particle and higher-order correlations in a dilute gas lets us represent evolution over the time scale $\tau\ll \Delta t$ in terms of a certain two-body propagator, which is effectively a finite-time generalization of the S-matrix from the infinite-time scattering problem of textbooks. This transition from the many- to the two-body problem is an essential first step in our larger discussion, but by way of additional contribution we also choose to make this transition from a somewhat unusual perspective. Instead of directly following the motions of actual particles themselves, we consider the evolution of the second-quantized destruction operator in position space. This formalism will allow us to make some key steps with more mathematical rigor, yet it can still be given an intuitive interpretation as a sort of \textit{It's a Wonderful Life} story for particles: how would things have happened differently if a particle had been removed at some earlier time?

\subsection{Second quantization}
The formalism of \emph{second quantization} describes quantum gases using operators $\hat{a}(\mathbf{r},t)$ and $\hat{a}^{\dagger}(\mathbf{r},t)$ which, respectively, remove and create a particle at position $\mathbf{r}$ and time $t$. The primary application of this formalism is dealing with quantum statistics, since the requirements of symmetrizing or anti-symmetrizing wave functions for bosons or fermions are automatically met by imposing the canonical (anti-)commutation relations
\begin{equation}\label{CCR}
\hat{a}(\mathbf{r},t)\hat{a}^{\dagger}(\mathbf{r}^{\prime},t)\pm \hat{a}^{\dagger}(\mathbf{r}^{\prime},t)\hat{a}(\mathbf{r},t) = \delta^{3}(\mathbf{r}-\mathbf{r}^{\prime})
\end{equation}
where the $+$ and $-$ signs apply to fermions and boson, respectively. Quantum statistics will be less important here, because we will not consider degenerate gases, but we will use the second-quantized formalism as a compactly written formalism for describing a system of many particles with pair-wise interactions. 

Letting the general mixed quantum state of the gas be described by the density operator $\hat{\rho}$, which acts in the enormously large many-body Hilbert space, the single-particle density matrix can be expressed in terms of $\hat{a}$ and $\hat{a}^\dagger$ operators as
\begin{eqnarray}\label{R1}
\rho_1(\mathbf{r},\mathbf{r}',t)&=&\frac{1}{N}\langle\hat{a}^{\dagger}(\mathbf{r}',t)\hat{a}(\mathbf{r},t)\rangle\;,
\end{eqnarray}
where $N$ is the total number of particles in the gas, and we define the expectation value of any operator $\hat{A}$ as $\langle\hat{A}\rangle := \mathrm{Tr}(\hat{\rho}\hat{A})$. Most accessible observables for the gas can then be extracted from $\rho_{1}$ without further reference to the full $\hat{\rho}$. The local gas density, for example, is $n(\mathbf{r})=N\rho_{1}(\mathbf{r},\mathbf{r})$; the momentum distribution is obtained by Fourier transforming $\rho_{1}$ with respect to both $\mathbf{r}$ and $\mathbf{r}'$. 

The goal of this paper will be an equation of motion for $\rho_{1}(\mathbf{r},\mathbf{r}',t)$, comparable to the classical Boltzmann equation, focusing on its description of the time evolution of quantum coherence. It will therefore be useful to have a way to compare our later results to the classical Boltzmann equation directly. For this purpose a close analog to the classical phase space distribution $f(\mathbf{r},\mathbf{p},t)$ is supplied by the Wigner function
\begin{eqnarray}\label{Wigner}
W(\mathbf{r},\mathbf{p},t) =\int\! \frac{d^{3}z}{(2\pi\hbar)^3}e^{i\mathbf{p}\cdot\mathbf{z}/\hbar}\rho_1\left(\mathbf{r}+\frac{\mathbf{z}}{2},\mathbf{r}-\frac{\mathbf{z}}{2},t\right)
\end{eqnarray}
which is simply a partial Fourier transform that presents the same information as $\rho_{1}$. By construction $W$ is a \emph{real} function in phase space, and furthermore it allows computation of quantum expectation values as integrals in phase space weighted by $W$, just as one computes averages in classical statistical mechanics with weighting by $f$. On the other hand $W$ can take negative values, and obeys constraints related to the Heisenberg uncertainty relation which have no classical counterparts. The Wigner function represents well the classical features of quantum states that are close to classical but it has fewer advantages for non-classical states. 

To re-examine Boltzmannian scattering in the quantum regime we will need to evolve $\rho_1$ over a finite time of order $\tau$. To determine time evolution we therefore return from $\rho_1$ to the $\hat{a}$ and $\hat{a}^\dagger$ operators themselves.

\subsection{Quantum evolution in the interaction picture}
In the Heisenberg picture of quantum mechanics, the operators $\hat{a}$ and $\hat{a}^{\dagger}$ become time-dependent. The evolution of $\hat{a}(\mathbf{r},t)$ is given, like that of any operator, by the Heisenberg equation of motion
\begin{eqnarray}\label{Heis0}
	i\hbar\partial_{t}\hat{a}(\mathbf{r},t) = [\hat{a}(\mathbf{r},t),\hat{H}]\;,
\end{eqnarray}
where $\hat{H}$ is the Hamiltonian operator. Here we will assume the absence of any external potential varying on short enough length scales that it significantly affects two-particle scattering---that is, we set the external potential to zero, but anticipate that it will be self-consistent to put it back into our final equation, in a kind of local-density approximation, as long as it varies slowly enough in space. We therefore consider the second-quantized Hamiltonian
\begin{widetext}
\begin{eqnarray}\label{H1}
\hat H &=& \int\!d^{3}r\,\frac{\hbar^{2}}{2m}\mathbf{\nabla}\hat{a}^{\dagger}(\mathbf{r})\cdot\mathbf{\nabla}\hat{a}(\mathbf{r})
 + \frac{1}{2}\int d^3r_{1} d^3r_{2}\, U(\mathbf{r}_{1}-\mathbf{r}_{2}) \hat{a}^\dagger(\mathbf{r}_{1}) \hat{a}^\dagger(\mathbf{r}_{2}) \hat{a}(\mathbf{r}_{2}) \hat{a}(\mathbf{r}_{1})
\end{eqnarray}
where $U(\mathbf{r}_{1}-\mathbf{r}_{2})$ is the potential for a short-ranged central force between two particles located respectively at $\mathbf{r}_{1}$ and $\mathbf{r}_{2}$. Applying the canonical (anti-)commutation relation (\ref{CCR}), we obtain the Heisenberg equation of motion
\begin{eqnarray}\label{Heis1}
i\hbar\partial_t\hat{a}(\mathbf{r},t)\!\! &=& -\frac{\hbar^{2}}{2m}\nabla^{2}\hat{a}(\mathbf{r},t)+\int\!d^{3}r'\, U(\mathbf{r}-\mathbf{r}^{\prime})\hat{a}^{\dagger}(\mathbf{r}^{\prime},t)\hat{a}(\mathbf{r}^{\prime},t)\hat{a}(\mathbf{r},t)\;
\end{eqnarray}
which is valid for both bosons and fermions.

The first term in $\hat{H}$ merely describes ballistic motion of particles. It will simplify our problem to factor out this trivial part of the time evolution, by adopting the interaction picture for second-quantized operators and writing
\begin{eqnarray}\label{intpic}
\hat{a}(\mathbf{r},t) = \int\!d^3q\,G_1(\mathbf{r},\mathbf{q},t-t_0)\hat{a}_I(\mathbf{q},t)\;\Longleftrightarrow\;\hat{a}_{I}(\mathbf{r},t) = \int\!d^3q\,G_1^*(\mathbf{r},\mathbf{q},t-t_0)\hat{a}(\mathbf{q},t)\;,
\end{eqnarray}
where $G_1$ is the free Schr\"odinger propagator
\begin{equation}\label{G1}
G_{1}(\mathbf{r},\mathbf{q},t) = \left(\frac{m}{2\pi i \hbar t}\right)^{\frac{3}{2}}e^{\frac{i m}{2\hbar t}|\mathbf{r}-\mathbf{q}|^{2}}\;,
\end{equation}
which satisfies
\begin{eqnarray}\label{G1delta}
\int\!d^{3}q\,G_{1}^{*}(\mathbf{r}',\mathbf{q},t)G_{1}(\mathbf{r},\mathbf{q},t) = \delta^{3}(\mathbf{r}-\mathbf{r}') = \lim_{t\to0}G_{1}(\mathbf{r},\mathbf{r}',t)\;.
\end{eqnarray}
The interaction picture operator $\hat{a}_{I}(\mathbf{r},t)$ thereby exactly equals $\hat{a}(\mathbf{r},t_0)$ at the arbitrarily chosen time $t=t_0$, but has the free ballistic motion part of its evolution artificially removed. The evolution of $\hat{a}_{I}$ will therefore represent the quantum analog of classical scattering.

\subsection{Quantum scattering}
\subsubsection{Formal solution with a closed hierarchy}
Equation~(\ref{Heis1}) can be integrated formally, from any initial time $t_0$ to any final time $t$. The formal solution is somewhat cumbersome to write explicitly, so we introduce it first in a compressed notation to make clear its pattern:
\begin{equation}
\hat{a}_{I}(\mathbf{r},t) = \sum_n \int\!dq\, (\hat{a}^{\dagger})^n\, (\hat{a})^{n+1}\, L_{n+1}\;.
\end{equation}
That is, the expansion is a series of normally ordered terms with $n$ creation operators on the left and $n+1$ destruction operators on the right, with a set of time-dependent c-number kernels $L_n$. If we now write the solution out explicitly it reads
\begin{eqnarray}\label{Ansatz}
\hat{a}_{I}(\mathbf{r},t) & = & \hat{a}_{I}(\mathbf{r},t_{0})\\
&&\!\!\!\!\!\!\!\!\!\!\!\!\!\!\!\!\!\!\!\!\!\!\!\!\!\!\!\!\!\!\!\!+\sum_{n=1}^{\infty}  \int d^{3(n+1)}q_j d^{3n}q'_{j}\, L_{n+1}(\mathbf{r};\mathbf{q}'_{1},...,\mathbf{q}'_{n};\mathbf{q}_{0},\mathbf{q}_{1},...,\mathbf{q}_{n};t-t_{0})\hat{a}^\dagger(\mathbf{q}'_{1},t_{0})...\hat{a}^\dagger(\mathbf{q}'_{n},t_{0})\,\hat{a}(\mathbf{q}_n,t_{0})...\hat{a}(\mathbf{q}_1,t_{0})\hat{a}(\mathbf{q}_0,t_{0})\;.
\end{eqnarray}\end{widetext}
The exact solution takes this form, without any approximation or loss of generality, because the Heisenberg equation of motion (\ref{Heis1}) for $\hat{a}(t)$ always relates expressions having one more $\hat{a}$ than $\hat{a}^{\dagger}$ operator, and because any product of $\hat{a}$'s and $\hat{a}^{\dagger}$'s can be expressed exactly as a sum of ``normal-ordered'' terms, in which all $\hat{a}^{\dagger}$s are to the left of all $\hat{a}$'s. 

The Heisenberg equation (\ref{Heis1}) furthermore implies specific equations of motion, of first order in time, for all the kernels $L_n$. The initial condition that $\hat{a}_I(\mathbf{r},t_0)$ equals itself implies the initial condition on all the $L_{n>1}$: they must all vanish at $t=0$. With this initial condition the equations of motion determine all the $L_n$, because in contrast to the famous Bogoliubov-Born-Green-Kirkwood-Yvon (BBGKY) hierarchy of equations of motion for correlation functions, the hierarchy of equations for the coefficients in the operator expansion $L_{n}$ is \textit{closed}. The equation of motion for each $L_{n}$ in this operator expansion involves only $L_{m}$ for $m\leq n$, and so each $L_{n}$ may in principle be determined exactly, without having to solve the full $N$-body problem of all $N$ particles in the entire gas.

\subsubsection{``Time travel logic''}
This time evolution of the second-quantized destruction operator according to  (\ref{Ansatz}) will be the basis of our whole picture of dilute gas dynamics. It may be an unfamiliar kind of picture, because it is not at all like the classical picture of particles moving and colliding. Instead it is like a science fiction film about time travel. If we first consider $t<t_{0}$, then what the expansion (\ref{Ansatz}) of $\hat{a}(\mathbf{r},t)$ says is that removing a particle from point $\mathbf{r}$ at a time in the past is equivalent to doing the following more complicated operation in the present ($t_{0}$): remove that one particle from $\mathbf{r}$, and also move a lot of other particles to the places where they would have been, if they hadn't interacted with the particle at $\mathbf{r}$ in the time since that $t$ in the past. This is like the plot premise of every time travel or alternate history story: changing one simple thing in the past is equivalent to making many more changes in the present. 

In fact we are interested in $t>t_{0}$, and in this case what (\ref{Ansatz}) says is that removing a particle from $\mathbf{r}$ at a time $t>t_{0}$ is equivalent to removing that particle back at $t_{0}$---and simultaneously re-arranging the other particles in such a way that they will now reach the same state at $t$ as they would have, if the particle at $\mathbf{r}$ had not been removed until $t$. This suggests a new kind of science fiction story, in which time travellers have to go back and change many details, in order to ensure that only one specific thing ends up changed in the present. In fact, though, this new time-travel challenge would be nothing other than the challenge we often face in real life: to control many small details in the present in order to succeed in changing just one specific thing in the future, without any extra undesirable consequences. The time-travel-ish logic of second quantized time evolution is not so unintuitive after all, once one is used to it. The destruction operator picture may be a weird way of describing the history of an interacting gas, but it does represent the information, and it has the advantage that the kernels $L_{n}$ are determined exactly by a hierarchy of equations that is closed for each $n$.

\subsubsection{Closure up to $n=2$}
For $n>2$, this determination is unfortunately only a determination-in-principle, because although solving the equation for each $L_{n}$ does not require solving the whole quantum $N$-body problem, it does require solving the quantum $n$-body problem, and even for $n=3$ this is usually too hard to achieve. Fortunately, however, $L_{2}$ can be obtained exactly, and in sufficiently explicit form to be useful. In Appendix \ref{Appendix:FromL2toT} we show that we can express ${L}_{2}$ as
\begin{eqnarray}\label{Teq0}
{L}_{2}(\mathbf{r},\mathbf{q}'_{1};\mathbf{q}_{2},\mathbf{q}_{1};t)&=&i\delta^{3}\left(\frac{\mathbf{r}+\mathbf{q}'_{1}}{2}-\frac{\mathbf{q}_{1}+\mathbf{q}_{2}}{2}\right)\nonumber\\
&&\times T(\mathbf{q}_{2}-\mathbf{q}_{1},\mathbf{r}-\mathbf{q}'_{1},t)\;,\end{eqnarray}
where the reduced kernel $T(\mathbf{s},\mathbf{s}',t)$ is a finite-time generalization of the T-matrix, in the position representation, for the effective one-body problem of a particle with reduced mass $m/2$ scattering from the central potential $U(|\mathbf{s}|)$. We will define $T(\mathbf{s},\mathbf{s}',t)$ explicitly in Section IV, below, and see that it describes scattered partial waves expanding spherically at finite speed behind a dispersively spreading wavefront. 

Although we thus have $L_{2}$ explicitly and exactly, we will not be able to compute the higher $L_{n>2}$, and in the end we will simply ignore them by appealing heuristically to low gas density. Our ultimate conclusions may thus not actually be any more rigorous than the same conclusions are when they are reached by truncating the BBGKY hierarchy. Inasmuch as we do not need to make the truncation approximation first in order to obtain closed equations, however, our approach is at least somewhat cleaner than the standard BBGKY truncation. 

\begin{widetext}
If we now observe both (\ref{Ansatz}) and its Hermitian conjugate, and apply the canonical (anti-)commutation relation to re-express all terms in normal order, we find
\begin{eqnarray}\label{R1op}
\hat{a}_I^{\dagger}(\mathbf{r}',t_{0}+\tau)\hat{a}_I(\mathbf{r},t_{0}+\tau) &=&\hat{a}^{\dagger}(\mathbf{q}',t_{0})\hat{a}(\mathbf{q},t_{0})\nonumber\\
&&+\int\,d^{6}q_{j}\,d^{6}q'_{j}\,K(\mathbf{r},\mathbf{r}';\mathbf{q}_{1},\mathbf{q}_{2}; \mathbf{q}'_{1},\mathbf{q}'_{2};\tau)\hat{a}^{\dagger}(\mathbf{q}'_{1},t_{0}) \hat{a}^{\dagger}(\mathbf{q}'_{2},t_{0})\hat{a}(\mathbf{q}_{2},t_{0})\hat{a}(\mathbf{q}_{1},t_{0})+\dots 
\nonumber\\
K(\mathbf{r},\mathbf{r}';\mathbf{q}_{1},\mathbf{q}_{2}; \mathbf{q}'_{1},\mathbf{q}'_{2};t)&=&\delta^3(\mathbf{r}'-\mathbf{q}'_{1}) 
{L}_{2}(\mathbf{r};\mathbf{q}'_{2};\mathbf{q}_{1},\mathbf{q}_{2};t)+\delta^3(\mathbf{r}-\mathbf{q}_{1}) {L}_{2}^{*}(\mathbf{r}',\mathbf{q}_{2};\mathbf{q}'_{1},\mathbf{q}'_{2};t)\nonumber\\
&&+\int\,d^{3}z\,{L}_{2}(\mathbf{r},\mathbf{z};\mathbf{q}_{1},\mathbf{q}_{2};t){L}_{2}^{*}(\mathbf{r}',\mathbf{z};\mathbf{q}'_{1},\mathbf{q}'_{2};t)
\;,\label{Kdef}
\end{eqnarray}
where $\dots$ indicates normally ordered terms with three or more $\hat{a}^{\dagger}$ operators to the left of three or more $\hat{a}$ operators. The fact that even in the expansion of the product $\hat{a}_I^\dagger\hat{a}_I$ the $L_{n\geq3}$ kernels still only appear in normally ordered terms with $n\geq3$ creation and destruction operators is another convenient exact hierarchy closure that we obtain in this operator formalism. When we follow Boltzmann by discarding the $n>2$ terms, the exactly defined kernel $L_2$ will be the only remaining element in our description of quantum gas dynamics.

\subsection{Diluteness and the \textit{Stosszahlansatz}}
We assume that our gas is (a) dilute and (b) not quantum degenerate. Accordingly we will ignore the possibilities of having multiple particles either within the same very small region of physical space, or within the same quantum orbital, on the grounds that these possibilities have no significant effect within the time frame of order $\tau$ that we consider. What this concretely means is that we will simply discard the $\dots$ terms in (\ref{R1op}). 

Furthermore, we will adopt the standard quantum statistical version of Boltzmann's heuristic replacement, the \textit{Stosszahlansatz} factorization:
\begin{eqnarray}\label{SZA}
\frac{1}{N(N-1)}\big\langle\hat{a}^{\dagger}(\mathbf{r}'_{1}) \hat{a}^{\dagger}(\mathbf{r}'_{2})\hat{a}(\mathbf{r}_{2})\hat{a}(\mathbf{r}_{1})\big\rangle\to \frac{1}{N^2}\Bigl( \big\langle\hat{a}^{\dagger}(\mathbf{r}'_{1}) \hat{a}(\mathbf{r}_{1})\big\rangle\,\big\langle\hat{a}^{\dagger}(\mathbf{r}'_{2})\hat{a}(\mathbf{r}_{2})\big\rangle
\pm \big\langle\hat{a}^{\dagger}(\mathbf{r}'_{1}) \hat{a}(\mathbf{r}_{2})\big\rangle\,\big\langle\hat{a}^{\dagger}(\mathbf{r}'_{2})\hat{a}(\mathbf{r}_{1})\big\rangle\Bigr)
\;,\end{eqnarray}
where the $+$ and $-$ alternatives apply for bosons and fermions, respectively. We will assume that this factorization holds at some given time $t_{0}$, and then study the later time evolution of the system to the time $t_{0}+\tau$. In fact the factorization ansatz cannot describe particles that are actually in the process of colliding at $t_{0}$, but we will assume that these initially interacting cases involve only a negligible fraction of the gas particles. The interactions that occur after $t_{0}$ will induce correlations, but we will only compute the evolution of the single-particle density matrix $\rho_1(\mathbf{r},\mathbf{r}',t)$. 

Finally, then, we will simply assume that our results will hold for \textit{any} time $t_{0}\to t$, even though it would seem that there can only be one time at which the factorization holds, because interactions induce correlations. 

We realize full well that all three of the approximations and assumptions just outlined are hard to justify rigorously, however reasonable they may seem. We assume them without further comment because we have nothing to add, in this paper, to their derivation. We focus instead on the issue that still remains after them: the approximation of finite-duration wave-mechanical collisions as instantaneous point-like events. 

When we make the heuristic dilute gas approximation of discarding the $\dots$ terms of (\ref{Kdef}), insert the reduction of ${L}_{2}(\mathbf{r},\mathbf{q}'_{1};\mathbf{q}_{2},\mathbf{q}_{1};t)$ to $T(\mathbf{q}_{2}-\mathbf{q}_{1},\mathbf{r}-\mathbf{q}'_{1},t)$ according to (\ref{Teq0}), and then take expectation values in an arbitrary many-body quantum state and apply the \textit{Stosszahlansatz} factorization (\ref{SZA}), we obtain for the single-particle density matrix in the interaction picture $\rho_{I}(\mathbf{r},\mathbf{r}',t)\equiv\langle\hat{a}_{I}^{\dagger}(\mathbf{r}',t)\hat{a}_{I}(\mathbf{r},t)\rangle$
\begin{eqnarray}\label{R1T}
\rho_{I}(\mathbf{r},\mathbf{r}',t_{0}+\tau)&=& \rho_{I}(\mathbf{r},\mathbf{r}',t_{0}) + 2(N-1)\Bigl(i\,R(\mathbf{r},\mathbf{r}',\tau,t_{0}) -i\, R^{*}(\mathbf{r}',\mathbf{r},\tau,t_{0})+S(\mathbf{r},\mathbf{r}',\tau,t_{0})\Bigr) \nonumber\\
R(\mathbf{r},\mathbf{r}',\tau,t_{0})&=&\int\!d^{3}s_{1}d^{3}s_{2}\,\tilde{T}(\mathbf{s}_{1},\mathbf{s}_{2},\tau)\rho_{I}\left(\mathbf{r}-\frac{\mathbf{s}_{1}-\mathbf{s}_{2}}{2},\mathbf{r}',t_{0}\right)\rho_{I}\left(\mathbf{r}-\frac{\mathbf{s}_{1}+\mathbf{s}_{2}}{2},\mathbf{r}-\mathbf{s}_{1},t_{0}\right)\nonumber\\
S(\mathbf{r},\mathbf{r}',\tau,t_{0})&=&\int\!d^{3}s_{1}d^{3}s_{2}d^{3}s'_{1}d^{3}s'_{2}\,\delta^{3}(\mathbf{r}-\mathbf{r}'-\mathbf{s}_{1}+\mathbf{s}'_{1})\tilde{T}(\mathbf{s}_{1},\mathbf{s}_{2},\tau)\tilde{T}^{*}(\mathbf{s}'_{1},\mathbf{s}'_{2},\tau)\nonumber\\
&&\qquad\qquad\qquad\qquad\times \rho_{I}\left(\mathbf{r}-\frac{\mathbf{s}_{1}-\mathbf{s}_{2}}{2},\mathbf{r}'-\frac{\mathbf{s}'_{1}-\mathbf{s}'_{2}}{2},t_{0}\right)\rho_{I}\left(\mathbf{r}-\frac{\mathbf{s}_{1}+\mathbf{s}_{2}}{2},\mathbf{r}'-\frac{\mathbf{s}'_{1}+\mathbf{s}'_{2}}{2},t_{0}\right)\nonumber\\
\tilde{T}(\mathbf{s},\mathbf{s}',\tau)&\equiv& \frac{T(\mathbf{s},\mathbf{s}',\tau)\pm T(\mathbf{s},-\mathbf{s}',\tau)}{2}
\;,
\end{eqnarray}
where the $\pm$ applies to bosons ($+$) and fermions ($-$). 

If all our particles were distinguishable instead of being quantum mechanically indistinguishable, we would have obtained (\ref{R1T}) with only a factor $(N-1)$ instead of $2(N-1)$ in the first line, and with simply $T$ itself instead of $\tilde{T}$ in the definitions of $R$ and $S$. We will soon see in Section IV, below, that $\tilde{T}$ is exactly $T$ except with the scattering amplitudes $f_{l}(k)$ set to zero for all odd $l$ (bosons) or all even $l$ (fermions). Apart from quantum degeneracy (which we do not consider), therefore, the effects of quantum statistics on scattering are simply (a) an overall factor of 2 in the collision terms for both fermions and bosons, and (b) an effective parity condition on scattering amplitudes, depending on particle statistics.

With Eqns.~(\ref{R1T}) we have achieved our first goal: we have implemented Boltzmann's heuristic \textit{Stosszahlansatz} approximation to obtain a closed equation for the time evolution of the single-particle density matrix, while not yet making any approximations whatever to quantum mechanical two-body scattering. In (\ref{R1T}), wave mechanical scattering is thus treated exactly, as a process that extends over finite range and duration. We are now in a position to investigate time evolution of quantum gases arbitrarily far from equilibrium, in the sense of having arbitrary single-particle density matrices, still within the heuristic Boltzmannian framework of neglecting higher-order correlations, but without imposing classical concepts of scattering onto quantum dynamics.

Eqns.~(\ref{R1T}) define evolution from an arbitrarily chosen initial time $t_0$ over the finite time $\tau$. To obtain any kind of Boltzmann-like equation from (\ref{R1T}) therefore requires further steps to extract a differential equation in time for $\rho_1(\mathbf{r},\mathbf{r}',t)$. We can anticipate that the $S$ term on the right-hand side of (\ref{R1T}), which involves a product of two of the finite-time T-matrix kernels $T$, will correspond to the classical Boltzmann $\Gamma_+$ term: as a product of two scattering amplitudes, it defines a probability density for particles to scatter into given final states. Together the two $R$ terms in (\ref{R1T}) will turn out to include a real part which provides a mean field potential effect like that of the Vlasov term in (\ref{BVE}), but also an imaginary part which is related, according to the optical theorem of quantum scattering, to the total scattering cross section. The imaginary part of the $R$ terms in (\ref{R1T}) will thus correspond to the classical Boltzmann $\Gamma_-$ term, and represent probability loss from initial states due to particles being scattered out of them.

These effects of quantum scattering will only correspond to their classical counterparts for sufficiently classical gas states $\rho_1$, however. The meaning of (\ref{R1T}) in general depends on the general form of the finite-time T-matrix $T$ as a representation of wave-mechanical scattering.

\section{Quantum Boltzmann equations}
\subsection{Quantum scattering under $T(\mathbf{s},\mathbf{s}',t)$}
In Appendix \ref{Appendix:FromL2toT} we show that we can express the finite-time T-matrix $T$ in (\ref{R1T}) as
\begin{eqnarray}\label{Teq}
T(\mathbf{s},\mathbf{s}',t)&=&\left(\frac{im}{4\pi\hbar t}\right)^{\frac{3}{2}}\int\!d^3\,s''\,e^{-i\frac{m}{4\hbar t}|\mathbf{s}-\mathbf{s}''|^2}\sum_{l=0}^{\infty}\frac{(2l+1)P_{l}(\hat{\mathbf{s}}''\cdot\hat{\mathbf{s}}')}{(2\pi)^{2}}\int_{-\infty}^{\infty}\!k^{3}dk\,e^{-i\frac{\hbar}{m}k^{2}t}f_{l}(k)h_{l}(ks'')h_{l}(ks')\;.
\end{eqnarray}
Here the $P_{l}(x)$ are Legendre polynomials, $h_{l}(r)$ are spherical Hankel functions, and $f_{l}(k)$ is the wavenumber-dependent scattering amplitude for the $l$-th partial wave. We draw attention to the lower limit of the integration over $k$: this really is $-\infty$, not zero, even though $k$ is a radial coordinate of the kind that is usually positive. See our Appendix for the detailed derivation, which has some non-trivial steps. 

We emphasize that (\ref{Teq}) is neither perturbative nor heuristic. Our expression for $T(\mathbf{s},\mathbf{s}',t)$ is valid exactly for all $|\mathbf{s}|$ and $|\mathbf{s}'|$ larger than the range of the potential $U$. To describe evolution within that short range, where two particles are actually interacting directly, our expression for $L_{2}$ must be supplemented with short-range information. The scattering amplitudes $f_{l}(k)$, however, are by definition the coefficients which exactly describe the longer range and longer term effects of the short-ranged potential $U$. They show correctly where the particles go, and when, after they have interacted; conversely they show where the particles must have been, before interacting, in order to have reached given points afterwards. The $f_{l}(k)$ are complex numbers which must be computed for any given $U(r)$ by solving the one-body Schr\"odinger equation; each is derived from a real \textit{phase shift} $\delta_{l}(k)$ according to 
\begin{equation}\label{optical}
f_{l}=\frac{e^{i\delta_{l}}\sin\delta_{l}}{k}\;,\qquad \hbox{which implies}\qquad\mathrm{Im}[f_{l}] \equiv k\,|f_{l}|^{2}
\end{equation}
identically for every $l$ independently. This is known as the Optical Theorem for partial wave scattering amplitudes. 

\subsubsection{Quantum statistics}
Equation (\ref{Teq}) is obtained by solving the effective \emph{single-particle} evolution problem for the relative coordinate between two particles. As such it includes no effects of quantum statistics. The difference between fermions and bosons arises in Eqn.~(\ref{R1T}), where what appears in the time evolution of the single-particle density matrix of the gas is not $T$ itself but rather $\tilde{T}(\mathbf{s},\mathbf{s}',\tau)=[T(\mathbf{s},\mathbf{s}',\tau)\pm T(\mathbf{s},-\mathbf{s}',\tau)]/2$, with $+$ for bosons and $-$ for fermions. The only place where the sign of the $\mathbf{s}'$ argument makes any difference in (\ref{Teq}) is in the argument of the Legendre polynomial: $P_{l}(-\hat{\mathbf{s}}''\cdot\hat{\mathbf{s}}')\equiv (-1)^{l}P_{l}(\hat{\mathbf{s}}''\cdot\hat{\mathbf{s}}')$. Thus $\tilde{T}$ is identical to $T$ with
\begin{equation}\label{tildef}
f_{l}(k)\longrightarrow \tilde{f}_{l}(k) \equiv \frac{1\pm (-1)^{l}}{2}f_{l}(k)\;,
\end{equation}
which is simply $f_{l}(k)$ with all its values for odd $l$ set to zero, for bosons, or for fermions, with all its values for even $l$ set to zero. In (\ref{R1T}) above we have already included a factor of two that is due to quantum statistics for both fermions and bosons; we have now shown what we mentioned above in Section III, that an effective parity condition on $f_{l}(k)$ is quantum statistics' only other effect on two-body scattering.

From now on we will in any case simply assume that the $f_l(k)$ have been obtained and are known, for whatever interparticle potential applies to a particular gas. We will therefore also from now on simply write $T$ for $\tilde{T}$ and $f_{l}$ for $\tilde{f}_{l}$, since as far as all our further results will be concerned, the difference between fermions and bosons is simply that the $f_{l}(k)$ of one parity or the other all happen to vanish.

\subsubsection{Secular component}
With the $f_{l}(k)$ thus given, the importance of (\ref{Teq}) is that it describes quantum mechanical scattering as a finite time evolution of waves propagating over finite distances. It will let us connect the formal quantum scattering problem, with its energy eigenstates that remain eternally time-independent and extend to infinity, to the Boltzmannian idealization of scattering as an instantaneous event at a point. We will be able to make this connection because we evolve in (\ref{R1T}) over the intermediate time scale $\tau$ which is much longer than the collision duration $\delta t$ and yet much shorter than the inverse collision rate $\Delta t$. We can therefore expand to leading order in the time scale ratio $\epsilon = \delta t/\tau = \tau/\Delta t$, so that $\tau$ will be infinite compared to $\delta t$ but infinitesimal compared to $\Delta t$. Secular (\textit{i.e.} steady, long-term) features of two-body scattering will thus yield an instantaneous differential equation in time for the density matrix $\rho_1(\mathbf{r},\mathbf{r}',t)$ of the gas as a whole.

Even with the insight about the time scale hierarchy, however, it appears \textit{not} to be possible in general to reduce the finite-time evolution equation (\ref{R1T}) with non-local $T(\mathbf{s},\mathbf{s}',\tau)$ kernels to a differential equation in time. That is, the reduction of (\ref{R1T}) to anything like a Boltzmann equation is only possible in cases where further assumptions about $\rho_{I}$ can be made. To see where the problem lies, it suffices to examine the Fourier transform of $T(\mathbf{s},\mathbf{s}',\tau)$ with respect to either one of its spatial arguments, for example $\mathbf{s}'$. After applying the relations
\begin{align}
e^{i\mathbf{k}\cdot{r}}&=\sum_{l=0}^{\infty}i^{l}(2l+1)j_{l}(kr)P_{l}(\mathbf{\hat{k}}\cdot\mathbf{\hat{r}})\label{planewave}\\
\oint\!d^{2}\mathbf{\hat{k}}\,P_{l}(\mathbf{\hat{k}}\cdot\mathbf{\hat{r}})P_{m}(\mathbf{\hat{k}}\cdot\mathbf{\hat{r}}')&=\frac{4\pi\delta_{lm}}{2l+1}P_{l}(\mathbf{\hat{r}}\cdot\mathbf{\hat{r}}')\label{Plint}\\
\int_{0}^{\infty}\!dr\,r^{2}h_{l}(kr)j_{l}(k'r)& =\lim_{\epsilon\to0^{+}} \frac{i}{k}\left(\frac{k'}{k}\right)^{l}\frac{1}{(k+i\epsilon)^{2}-k^{2}}\label{jhint}
\end{align}
where $j_{l}$ is the spherical Bessel function of order $l$, we obtain
\begin{align}\label{FTT}
\int\!d^{3}s'\,e^{i\mathbf{k}_{0}\cdot\mathbf{s}'}T(\mathbf{s},\mathbf{s}',t) &=\left(\frac{im}{4\pi\hbar t}\right)^{\frac{3}{2}}\int\!d^3\,s''\,e^{-i\frac{m}{4\hbar t}|\mathbf{s}-\mathbf{s}''|^2}\nonumber\\
&\times\sum_{l=0}^{\infty}i^{(l+1)}\frac{(2l+1)P_{l}(\hat{\mathbf{s}}''\cdot\hat{\mathbf{k}}_{0})}{\pi k_{0}}\int_{-\infty}^{\infty}\!k^{3}dk\,e^{-i\frac{\hbar}{m}k^{2}t}\frac{f_{l}(k)\left(\frac{k_{0}}{k}\right)^{l}h_{l}(ks'')}{(k+i\epsilon)^{2}-k_{0}^{2}}\;.
\end{align}

We must now distinguish the secular component in (\ref{FTT}), representing the long-term consequences of collisions as completed events, from the transient component, which describes short-term behavior during collisions as ongoing processes. We can do so by evaluating the $k$ integral in (\ref{FTT}) using the method of stationary phase, by deforming the contour of integration over $k$ in  into the plane of complex $k$, in such a way that the $e^{-i\frac{\hbar}{m}k^{2}t}$ factor in the $k$ integrand becomes a Gaussian that becomes narrower with increasing $t$. In doing this we must take note of the fact that the spherical Hankel function $h_{l}(ks'')$ becomes a rapidly oscillating function of $k$ when $s''$ is large, and that since we are integrating over all $s''$, large values of $s''$ must be considered. We can correctly incorporate this highly oscillator behavior of $h_{l}$ in our stationary phase integration by including the asymptotic form of the Hankel function
\begin{equation}
\lim_{kr\to\infty}h_{l}(kr) =i^{-(l+1)}\frac{e^{ikr}}{kr}\;,\label{limhank}
\end{equation}
and thereby seeing that the saddlepoint of the rapidly varying exponent in the $k$ integrand is at $k_{sp}=ms''/(2\hbar t)$. 

The stationary phase contour which passes through $k_{sp}$ thus may or may not have to detour around the integrand pole at $k = k_{0}=|\mathbf{k}_{0}|$, depending on whether $s"$ is larger or smaller than $2\hbar k_{0}t/m$. (Recall that $m/2$ is the reduced mass of the relative coordinate of two colliding particles of mass $m$.) See Fig.~\ref{polesfig}. 
\begin{figure}\label{polesfig}
\begin{center}
\includegraphics[width=0.4\textwidth]{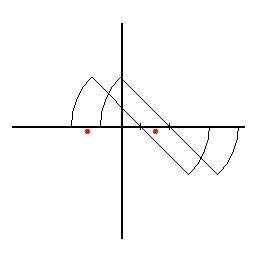}
\end{center}\caption{Time-dependent contour of integration to evaluate the integral (\ref{FTT}) in the saddlepoint approximation. For any given distance $s$ there is a time $2\hbar k_{0}t/m=s$ when the path crosses the saddlepoint, leaving it outside the contour.}
\end{figure}
Since the whole contribution to the $k$ integral in (\ref{FTT}) from the diagonal path through the saddlepoint is transient, in the sense that it decays with increasing $t$, and since the contribution from the counter-clockwise circled pole is given by Cauchy's theorem, we find
\begin{align}\label{FTT2}
\int\!d^{3}s'\,e^{i\mathbf{k}_{0}\cdot\mathbf{s}'}T(\mathbf{s},\mathbf{s}',t) &=\mathrm{(transient)}-i\left(\frac{im}{4\pi\hbar t}\right)^{\frac{3}{2}}\int\!d^3\,s''\,e^{-i\frac{m}{4\hbar t}|\mathbf{s}-\mathbf{s}''|^2}\theta\left(2\frac{\hbar k_{0}}{m}t-s''\right)e^{-i\frac{\hbar}{m}k_{0}^{2}t}e^{ik_{0}s''}f(k_{0},\hat{\mathbf{s}}''\cdot\hat{\mathbf{k}}_{0})\;,
\end{align}
where $\theta(x)$ is the Heaviside step function and
\begin{equation}f(k,\cos\theta) = \sum_{l=0}^{\infty}(2l+1)f_{l}(k)P_{l}(\cos\theta)\end{equation}
is the scattering amplitude at the deflection angle $\theta$ and momentum transfer $k$. 

As we have said, the kernel $T(\mathbf{s},\mathbf{s}'',t)$ is a finite-time T-matrix. The integral over $\mathbf{s}''$ with the $e^{-i\frac{m}{4\hbar t}|\mathbf{s}-\mathbf{s}''|^2}$ factor in (\ref{FTT2}) is just the free-particle propagation that distinguishes the time-evolution operator $\hat{U}$ from the $\hat{S}$ matrix of the interaction picture. If we therefore define the finite-time \textit{scattered wave} $\psi_{\mathrm{scat}}(k_{0},\mathbf{r},t)$ in terms of $T$ as
\begin{align}
\int\!d^{3}s''\,[\delta^{3}(\mathbf{s}-\mathbf{s}')+iT(\mathbf{s},\mathbf{s}',t)]e^{i\mathbf{k}_{0}\cdot\mathbf{s}'} \equiv e^{-i\frac{\hbar}{m}k_{0}^{2}t}\left(\frac{im}{4\pi\hbar t}\right)^{\frac{3}{2}}\int\!d^3\,s''\,e^{-i\frac{m}{4\hbar t}|\mathbf{s}-\mathbf{s}''|^2}\left[e^{i\mathbf{k}_{0}\cdot\mathbf{s}''}+\psi_{\mathrm{scat}}(k_{0},\mathbf{s}'',t)\right]\;,
\end{align}
then according to (\ref{FTT2}) we have
\begin{align}\label{}
	\psi_{\mathrm{scat}}(k_{0},\mathbf{r},t) = \mathrm{(transient)}+\theta\left(2\frac{\hbar k_{0}}{m}t-r\right)\psi_{\mathrm{scat}}(k_{0},\mathbf{r})
\end{align}
where the time-independent $\psi_{\mathrm{scat}}(k_{0},\mathbf{r})$ is the scattered wave of textbook scattering theory.
Thus, apart from the transient term which we have suppressed, $T$ simply generates the scattered wave $\psi_{\mathrm{scat}}$ from textbook scattering theory, only within a finite sphere of expanding radius $2\hbar k_{0}t/m$ ($m/2$ again being the reduced mass of the relative coordinate of the two colliding particles). In the limit of infinite $t$ the transient vanishes and the scattered wave extends to infinite radius as in the textbooks. 

\begin{figure}
\begin{center}
\includegraphics[width=0.5\textwidth]{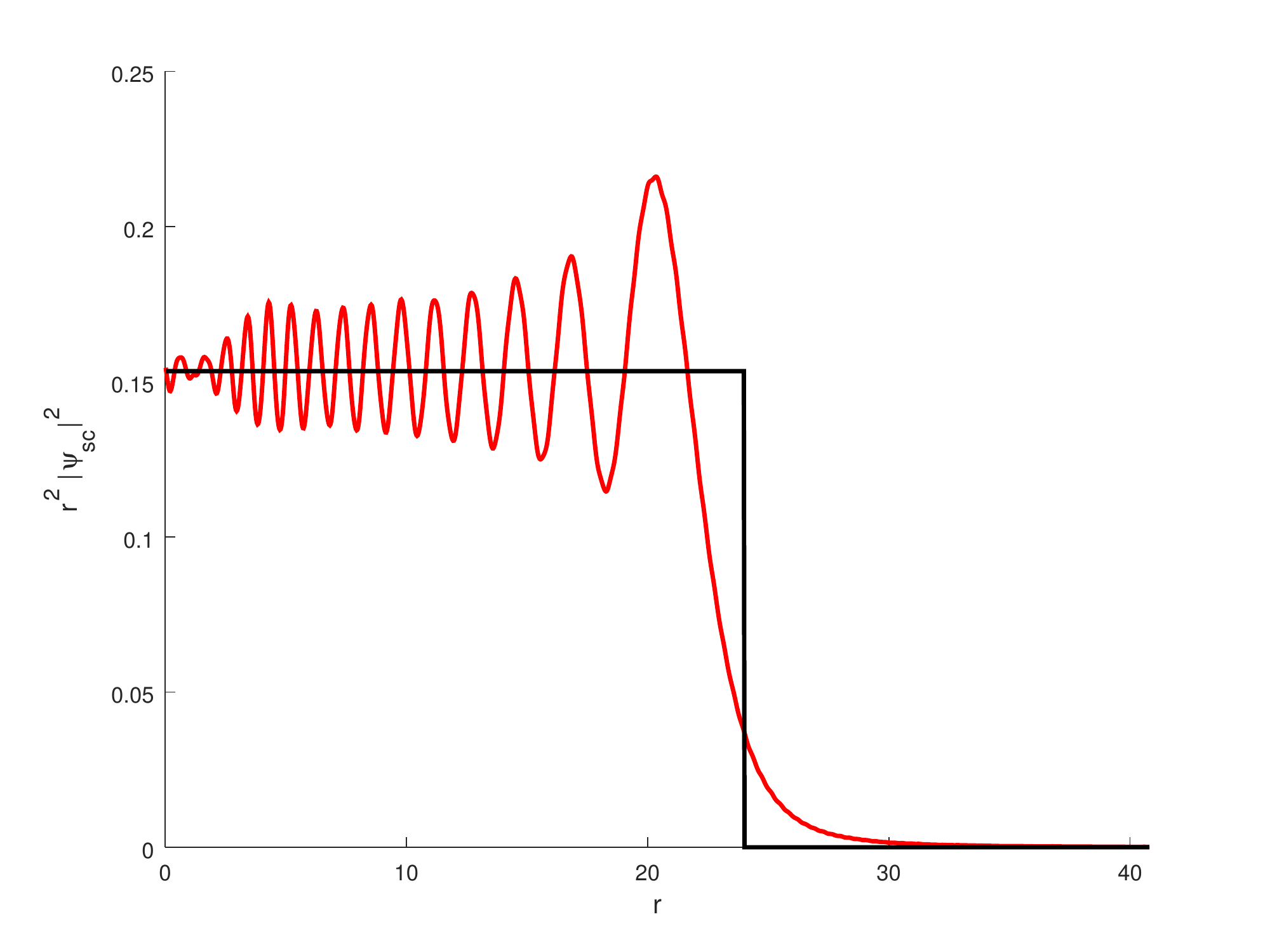}\caption{Comparison between the exact wave front density $|\psi_{scatt}(r)|^2 \,r^2$ (red) with the corresponding steepest-descent approximation (black). }\label{comparing_times_SD}
\end{center}
\end{figure}

At finite times, Fig.~(\ref{comparing_times_SD}) compares the secular component which is displayed in (\ref{FTT2}) to the exact result for $|\psi_{\mathrm{scat}}|^{2}$, for the simple $f_{l}(k)=a\delta_{l0}/(1+iak)$ of purely s-wave scattering. We see that the effects which we have suppressed as `transient' include small oscillations superimposed on the step-function form of the secular component, as well as dispersive broadening of the expanding wavefront. The spreading of the wavefront increases dispersively with $\sqrt{t}$, while the radius of the secular step function grows linearly in $t$. The significant secular effect of scattering is thus captured well by the stationary phase calculation that we have shown. And we have confirmed that our kernels $T(\mathbf{s},\mathbf{s}',t)$ do indeed represent quantum mechanical two-body scattering as a finite-duration process with long-term consequences.

At the same time we see the basic difficulty in deriving a quantum Boltzmann equation which will really be similar to the classical Boltzmann equation in describing scattering as an instantaneous local event. Equation (\ref{FTT2}) may look very much like a quantum analog to Boltzmannian scattering, but this is only because $\mathbf{k}_{0}$ was given as input: our Fourier transform of $T(\mathbf{s},\mathbf{s}',t)$ with respect to $\mathbf{s}'$ represents evolution from the special case of an initial plane wave for the relative coordinate of two colliding particles. In general, however, this relative coordinate does not have a definite wave number, or even a probabilistic mixture of different wave numbers, because the single-particle density matrix $\rho(\mathbf{r},\mathbf{r}')$ is in general not translation-invariant. We may say that the action of the kernel $T$, regarded as an operator, is to propagate each Fourier component of the initial relative-coordinate wave function as in (\ref{FTT2}). This means, however, that the operation of $T$ is essentially non-local, since it must parse the initial state of relative coordinate into Fourier components and this requires sampling a sufficient volume of space.

Even in classical physics, of course, collisions are not really instantaneous events that occur at a single point, because atoms and molecules have finite size and interactions between them have non-zero range. The Boltzmannian idealization of collisions as point-like events really only means that the spatial extent of a collision should be small compared to scales over which the collective properties of the gas can be observed to vary. The Heisenberg relation of quantum mechanics complicates this idealization with an additional issue that does not arise classically, however. Quantum mechanical scattering can only be idealized as point-like if the gas is homogeneous on a scale which may be much longer than the range of interactions, namely the wavelength associated with the typical relative momentum of nearby particles. This condition is not necessarily satisfied by all non-degenerate quantum gases, but it may be satisfied quite often in realistic cases. We will now show that when this condition is satisfied, one can derive the classical Boltzmann equation from our (\ref{R1T}), as an equation governing the Wigner function derived from the single-particle density matrix $\rho$ of the quantum gas.

\subsection{The classical Boltzmann equation for a classical gas}
We can express the single-particle Wigner function in terms of dimensionless variables by defining the two length scales $\lambda$ and $L$ which characterize its dependence on position and momentum:
\begin{eqnarray}\label{Wscales}
W(\mathbf{r},\mathbf{p},t) \equiv \tilde{W}(\mathbf{r}/L,\mathbf{p}\lambda/\hbar,t)
\Rightarrow\rho_{I}(\mathbf{r},\mathbf{r'},t) = \int\!d^{3}p\, e^{-i\mathbf{p}\cdot\frac{\mathbf{r}-\mathbf{r}'}{\hbar}}\tilde{W}\left(\frac{\mathbf{r}+\mathbf{r}'}{2L},\frac{\mathbf{p}\lambda}{\hbar}\right)\nonumber
\end{eqnarray}
where the assumption is that $\tilde{W}$ depends on its dimensionless arguments in a way that involves no very large or small numbers. 

This assumption about $\tilde{W}$ implies that the characteristic widths of $W$ in $\mathbf{r}$ and $\mathbf{p}$, respectively, are $\Delta r \sim L$ and $\Delta p\sim \hbar/\lambda$. In quantum terms we can identify $\lambda$ as a characteristic \textit{coherence length} for the gas, whereas $L$ represents the scale on which the gas state is spatially inhomogeneous (since spatial translation shifts all position arguments equally, leaving $\mathbf{r}-\mathbf{r}'$ unchanged). The homogeneity condition which is needed to treat quantum collisions as point-like then turns out to be the same as the classicality condition of being far from the Heisenberg limit: $\Delta r\Delta p\gg \hbar$ is equivalent to $L\gg\lambda$. Even when quantum statistics are not important, as we have assumed throughout this paper, a quantum gas may fail to be classical if it is in a non-equilibrium state that varies significantly on the scale of its local coherence length. We will consider below how the gas may behave in such states; for now we will stay in the classical limit by assuming $\lambda = \epsilon L$ for $\epsilon \ll 1$.

If we insert (\ref{Wscales}) into the right-hand side of (\ref{R1T}), while also rescaling the integration variables $\mathbf{p}\to\mathbf{p}/\lambda$ and $\mathbf{s}\to \mathbf{s} \lambda$, we find
\begin{eqnarray}\label{R1TWScales}
R(\mathbf{r},\mathbf{r}',\tau,t_{0})&=&\int\!d^{3}s_{1}d^{3}s_{2}d^3pd^3p^\prime\,e^{\frac{i}{2\hbar}(\mathbf{p}-\mathbf{p}^\prime)\cdot (\mathbf{s}_1-\mathbf{s}_2)} \,e^{-\frac{i}{\hbar}\mathbf{p}\cdot \frac{\mathbf{r}-\mathbf{r}^\prime}{\lambda}}\,T(\lambda\mathbf{s}_{1},\lambda{\mathbf{s}_{2}},\tau) 
\nonumber\\
&&\times \tilde{W}\left(\frac{\mathbf{r}+\mathbf{r}^\prime}{2L}-\epsilon \frac{\mathbf{s}_1-\mathbf{s}_2}{2},\mathbf{p},t_0\right)\,\tilde{W}\left(\frac{\mathbf{r}+\mathbf{r}'}{2L}+\epsilon\Big[\frac{\mathbf{r}-\mathbf{r}'}{2\lambda}-\frac{3\mathbf{s}_1-\mathbf{s}_2}{2}\Big],\mathbf{p}^\prime, t_0 \right)  \nonumber \\
S(\mathbf{r},\mathbf{r}',\tau,t_{0})&=&\int\!d^{3}s_{1}d^{3}s_{2}d^{3}s'_{1}d^{3}s'_{2}d^3pd^3p^\prime\,\delta^{3}(\mathbf{r}-\mathbf{r}'-\lambda\mathbf{s}_{1}+\lambda\mathbf{s}'_{1})\,e^{-\frac{i}{2\hbar}(\mathbf{p}-\mathbf{p}^\prime)\cdot (\mathbf{s}_2-\mathbf{s}_2^\prime)}\,e^{-\frac{i}{2\hbar} (\mathbf{p}+\mathbf{p}^\prime)\cdot\frac{\mathbf{r}-\mathbf{r}^\prime}{\lambda}}\nonumber\\
&&\times T(\lambda\mathbf{s}_{1},\lambda\mathbf{s}_{2},\tau) T^{*}(\lambda\mathbf{s}'_{1},\lambda\mathbf{s}'_{2},\tau) \nonumber\\
&&\times \tilde{W}\left(\frac{\mathbf{r}+\mathbf{r}^\prime}{2L}-\epsilon\frac{\mathbf{s}_1+\mathbf{s}_1^\prime -\mathbf{s}_2-\mathbf{s}_2^\prime}{2},\mathbf{p},t_0\right)\, \tilde{W}\left(\frac{\mathbf{r}+\mathbf{r}^\prime}{2L}-\epsilon\frac{\mathbf{s}_1+\mathbf{s}_1^\prime +\mathbf{s}_2+\mathbf{s}_2^\prime}{2},\mathbf{p}^\prime,t_0\right)  \;.
\end{eqnarray}
In the classical limit $\epsilon\to0$ (coherence length much shorter than the scale over which the gas properties vary), the dependence of the $\tilde{W}$ terms in the integrands on $\mathbf{s}_{1,2}$ and $\mathbf{s}'_{1,2}$ thus disappears from (\ref{R1TWScales}). Since the integrations over $\mathbf{p}$ and $\mathbf{p}'$ with the factor $e^{-\frac{i}{\hbar}\mathbf{p}\cdot\frac{\mathbf{r}-\mathbf{r}'}{\lambda}}$ also mean that $R$ is negligible unless $|\mathbf{r}-\mathbf{r}'|\lesssim \lambda \ll L$, the dependence of the second $\tilde{W}$ in $R$ on $\mathbf{r}-\mathbf{r}'$ can also be ignored to leading order in $\epsilon$. 

We can express (\ref{R1T}) itself in the Wigner representation (\ref{Wigner}), defining the interaction-picture single-particle Wigner function of the gas as
\begin{equation}\label{rho1Wig}
W_I(\mathbf{r},\mathbf{p},t_0+\tau) = \int\! \frac{d^{3}z}{(2\pi\hbar)^3}e^{i\mathbf{p}\cdot\mathbf{z}/\hbar}\rho_I\left(\mathbf{r}+\frac{\mathbf{z}}{2},\mathbf{r}-\frac{\mathbf{z}}{2},t_0+\tau\right)=W_{I0} + \epsilon W_{I1} + \dots
\end{equation}
with an expansion of $W_I$ in powers of $\epsilon = \lambda/L$. We then insert (\ref{R1T}) 
\begin{equation}
\rho_I(\mathbf{r},\mathbf{r}',t_0+\tau) = \rho_I(\mathbf{r},\mathbf{r}',t_0)+2(N-1)\left(iR(\mathbf{r},\mathbf{r}',t_0,\tau)-iR^*(\mathbf{r}',\mathbf{r},t_0,\tau)+S(\mathbf{r},\mathbf{r}',t_0,\tau)\right)\nonumber
\end{equation}
into (\ref{rho1Wig}) to obtain at zeroth order
\begin{align}
W_{I0}(\mathbf{r},\mathbf{p},t_0+\tau) =& W_{I}(\mathbf{r},\mathbf{p},t_0)\left[1 - 4(N-1) \int\!d^3p'\,W_I(\mathbf{r},\mathbf{p}',t_0)\mathrm{Im}[\mathcal{T}(\frac{\mathbf{p}-\mathbf{p}'}{2},\frac{\mathbf{p}-\mathbf{p}'}{2},\tau)]\right]\nonumber\\
& +2(N-1) \int\!\frac{d^3p'd^3p''}{(2\pi\hbar)^3}W_I(\mathbf{r},\mathbf{p}',t_0)W_I(\mathbf{r},\mathbf{p}'',t_0)\Big\vert\mathcal{T}(\mathbf{p}-\frac{\mathbf{p}'+\mathbf{p}''}{2},\frac{\mathbf{p}'-\mathbf{p}''}{2},\tau)\Big\vert^2\nonumber\\
\mathcal{T}(\mathbf{p},\mathbf{p}',\tau)\equiv&
\int\! d^3s_1 d^3s_2\,e^{\frac{i}{\hbar}(\mathbf{p}\cdot\mathbf{s}-\mathbf{p}'\cdot\mathbf{s}')}T(\mathbf{s},\mathbf{s}',\tau)\;.
\end{align}

\begin{eqnarray}\label{R1TWScalesLocalInPosition}
R(\mathbf{r},\mathbf{r}',\tau,t_{0})&=&\int\! d^3pd^3p^\prime\, e^{\frac{i}{\hbar}\mathbf{p}\cdot (\mathbf{r}-\mathbf{r}^\prime)}\, T_R\left(\frac{\mathbf{p}-\mathbf{p}^\prime}{2},\tau\right)\, W\left(\frac{\mathbf{r}+\mathbf{r}^\prime}{2},\mathbf{p},t_0\right)W\left(\mathbf{r},\mathbf{p}^\prime,t_0\right) \nonumber \\
T_R(\mathbf{q},\tau) &=& \int d^{3}s_{1}d^{3}s_{2}\, e^{-\frac{i}{\hbar}\mathbf{q}\cdot (\mathbf{s}_1-\mathbf{s}_2)}\,[T(\mathbf{s}_{1},\mathbf{s}_{2},\tau)\pm T(\mathbf{s}_{1},-\mathbf{s}_{2},\tau)] \\
S(\mathbf{r},\mathbf{r}',\tau,t_{0})&=&\int\! d^3pd^3p^\prime\,e^{-\frac{i}{\hbar} \frac{\mathbf{p}+\mathbf{p}^\prime}{2}\cdot(\mathbf{r}-\mathbf{r}^\prime)}\, T_S\left(\frac{\mathbf{p}-\mathbf{p}^\prime}{2},\mathbf{r}-\mathbf{r}^\prime,\tau\right)\, W\left(\frac{\mathbf{r}+\mathbf{r}^\prime}{2},\mathbf{p},t_0\right)\, W\left(\frac{\mathbf{r}+\mathbf{r}^\prime}{2},\mathbf{p}^\prime,t_0\right)\nonumber\\
T_S(\mathbf{q},\mathbf{z},\tau) &=& \int d^{3}s_{1}d^{3}s_{2}d^{3}s'_{1}d^{3}s'_{2}\, \delta^{3}(\mathbf{z}-\mathbf{s}_{1}+\mathbf{s}'_{1}) \,e^{-\frac{i}{\hbar}\mathbf{q}\cdot (\mathbf{s}_2-\mathbf{s}_2^\prime)}\, T(\mathbf{s}_{1},\mathbf{s}_{2},\tau)[T^{*}(\mathbf{s}'_{1},\mathbf{s}'_{2},\tau)\pm T^{*}(\mathbf{s}'_{1},-\mathbf{s}'_{2},\tau)] \;.
\end{eqnarray}

The step from (\ref{}) to (\ref{}) in the classical limit is decisive because it will allow us to reduce the temporal finite-difference equation (\ref{}) to a differential equation in time.

We can understand this step as the quantum version of Boltzmann's assumption mentioned before: that scattering is an event {\it local in position}. If the post-scattering reach, i.e. the distance covered by a particle after being scattered and flyiing a time $\tau$, is short enough, then it is not necessary to take into account density variations along its path (integration variables in the first argument of $W$) and it is only important to consider all different momentum components of the wave packet associated to it (second argument of $W$).\\
It is also interesting to point out what the physical meaning of the kernels $T_{R,S}$ is. In the interaction picture, the T-matrix $T(\mathbf{s_1},\mathbf{s}_2)$ allows to obtain by integration the scattered wave out of an incoming wave. The integral kernel $T_R(\mathbf{q},\tau)$ can be then understood as the overlapping between an incoming plane wave of momentum $\mathbf{q}$ and the corresponding scattered wave after a time $\tau$. In the case of $T_S(\mathbf{q},\mathbf{z},\tau)$, it is the overlapping between two scattered waves of the same momentum $\mathbf{q}$ when the scatterers are separated by a distance $\mathbf{z}$.\\

We will work first the terms in $R(\mathbf{r},\mathbf{r}',\tau,t_{0})$
\begin{eqnarray}\label{R0}
T_R(\mathbf{q},\tau)&=&\left(\frac{im}{4\pi\hbar \tau}\right)^{\frac{3}{2}}\sum_{l=0}^{\infty}\frac{(2l+1)[1\pm (-1)^{l}]}{(2\pi)^{2}}\int_{-\infty}^{\infty}\!k^{3}dk\,f_{l}(k)e^{-i\frac{\hbar \tau}{m}k^{2}}\nonumber\\
 & & \times \int\!d^{3}sd^{3}s'd^3s''\,
P_{l}(\hat{\mathbf{s}}''\cdot\hat{\mathbf{s}}')h_{l}(ks'')h_{l}(ks')\,  e^{-i\frac{m}{4\hbar \tau}|\mathbf{s}-\mathbf{s}''|^2}e^{\frac{i}{\hbar}\mathbf{q}\cdot(\mathbf{s}-\mathbf{s}'')} e^{-\frac{i}{\hbar}\mathbf{q} \cdot(\mathbf{s}''-\mathbf{s}')}\;.
\end{eqnarray}
The $[1\pm(-1)^{l}]$ factor comes from the $[T(\mathbf{s}_{1},\mathbf{s}_{2},\tau)\pm T(\mathbf{s}_{1},-\mathbf{s}_{2},\tau)]$ term in (\ref{R1T}), which in turn came from quantum statistics (Bose $=+$, Fermi $=1$) in the \textit{Stosszahlansatz} factorization (\ref{SZA}). This factor means, for example, that spin-polarized fermions have no s-wave scattering, and bosons have no p-wave. Quantum statistics has this effect on two-body scattering, even when there are no ideal gas effects of quantum statistics because the gas is far from quantum degeneracy. Since this factor is an inherent quantum statistical feature of scattering, we will simplify our notation from now on by absorbing it into the partial wave scattering amplitudes and writing $\tilde{f}_{l}(k)\equiv[1\pm(-1)^{l}]f_{l}(k)$.


We then use the expansion of a plane wave in spherical Bessel functions and Legendre polynomials
\begin{equation}
e^{i\mathbf{k}\cdot\mathbf{s}}\equiv\sum_{l=0}^{\infty}i^{l}(2l+1)j_{l}(ks)P_{l}(\hat{\mathbf{k}}\cdot\hat{\mathbf{s}})\;,
\end{equation}
the Legendre polynomials identities
\begin{eqnarray}
\oint\,d^{2}\hat{\mathbf{s}}\,P_{l}(\hat{\mathbf{k}}\cdot\hat{\mathbf{s}})P_{l'}(\hat{\mathbf{k}}'\cdot\hat{\mathbf{s}}) &\equiv& \frac{4\pi}{2l+1}\delta_{ll'}P_{l}(\hat{\mathbf{k}}\cdot\hat{\mathbf{k}}')\nonumber\\
P_{l}(1)&=&1\;,
\end{eqnarray}
and the Bessel and Hankel function integral
\begin{equation}
\int_{0}^{\infty}\!ds\,s^{2}\,j_{l}(k's)h_{l}(ks) = \frac{i}{k}\left(\frac{k'}{k}\right)^{l}\frac{1}{(k+i\epsilon)^{2}-k^{'2}}\qquad,\qquad \epsilon\to0^{+}
\end{equation}
to perform the $\mathbf{s}'$ and $\mathbf{s}''$ integrals, leaving
\begin{eqnarray}\label{R2}
T_R(\hbar \mathbf{k},\tau)&\doteq &-\sum_{l=0}^{\infty}\frac{(2l+1)}{\pi} \int_{-\infty}^{\infty}\!dk\,k^\prime\,\tilde{f}_{l}(k^\prime)\left(\frac{k}{k^\prime}\right)^{2l}\frac{e^{i\frac{\hbar \tau}{m}\left(k^{2}-k^{\prime 2}\right)}}{[(k^\prime +i\epsilon)^{2}-k^{2}]^{2}}
\;.
\end{eqnarray}
\end{widetext}

For a simpler notation we have rescaled the momentum argument into wavenumber $\mathbf{q}\rightarrow \hbar \mathbf{k}$ in the expression above.

Our entire goal is to express the effects of scattering which can cumulatively affect the long-term evolution of the gas, without having to mention the transient details of how pairs of particles affect each other during collisions. In wave mechanical scattering the wave function of the relative coordinate of two colliding particles only attains its asymptotic form at separations that are much larger than both the range of the interaction potential $U(r)$ and the de Broglie wavelength associated with the energy of the relative motion. This means that the duration of the quantum collision cannot be shorter than $\delta t = \lambda /v = m\lambda^{2}/\hbar$, and so our definition of the intermediate time scale $\tau\gg\delta t$ implies that
\begin{equation}\label{taulong}
\frac{\hbar\tau}{m\lambda^{2}}\gg 1\;.
\end{equation}
We can therefore identify the long-term effects of collisions by extracting the part of $T_R(\hbar\mathbf{k},\tau)$ which does not vanish for $\frac{\hbar\tau}{m\lambda^{2}}\to\infty$. Any components of $T_R(\hbar\mathbf{k},\tau)$ which decay for large $\tau$ can be neglected as transient behavior during the collision itself.
\begin{figure}
\begin{center}
\includegraphics[width=0.4\textwidth]{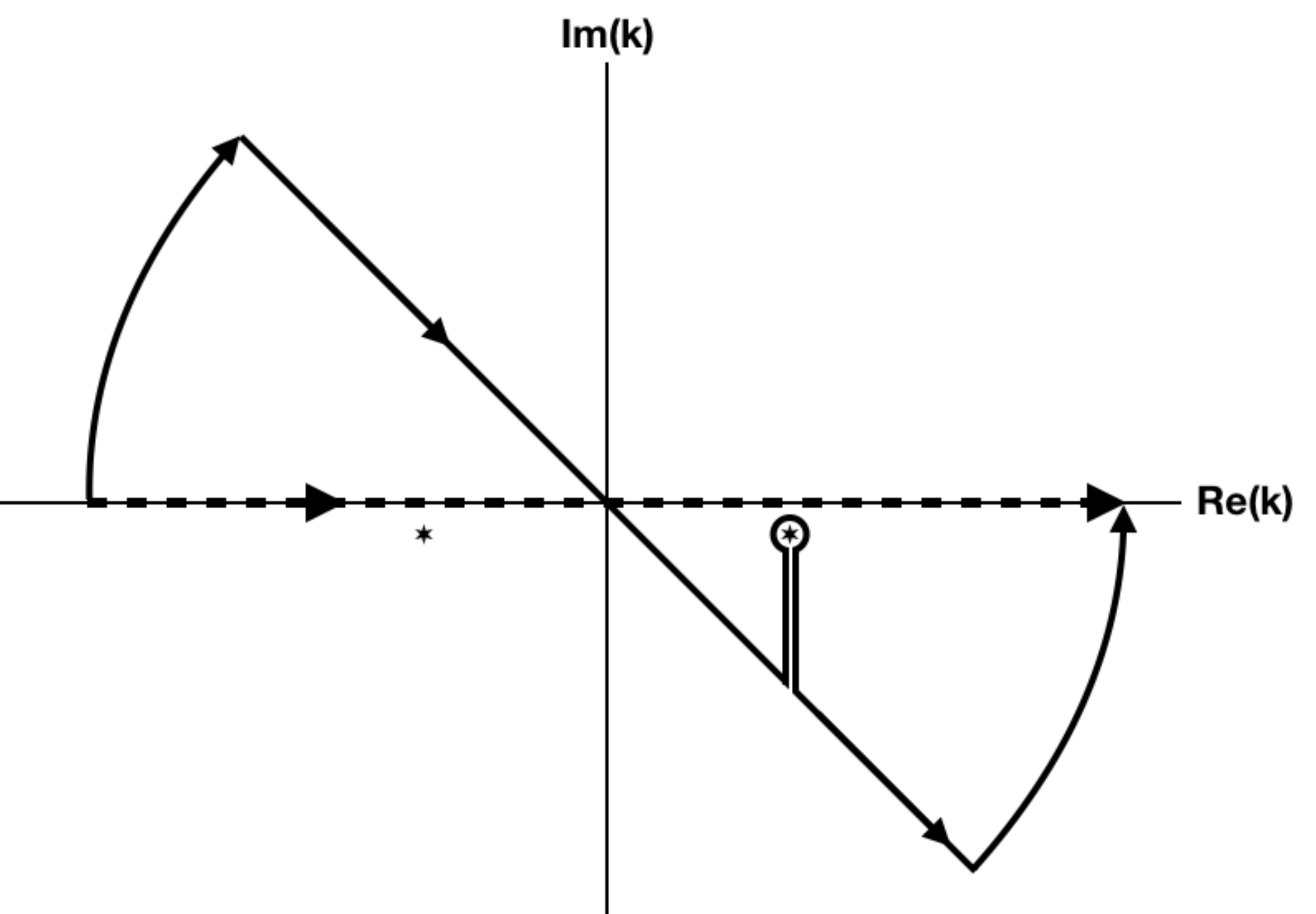}
\end{center}
\caption{Deformed (solid) and original (dashed, along real axis) contours of integration over $k$ in the kernel $T_R(\mathbf{k},\tau)$ of Eqn.~(\ref{R1T}). The two stars just below the real axis indicate the integrand singularities and only the one at $k^\prime=k-i\epsilon$ will yield a contribution of $O(\tau)$.}
\label{ContourFig}
\end{figure}

The way to distinguish secular and transient terms in $T_R(\hbar\mathbf{k},\tau,t_{0})$ is to deform the contour of integration over $k$ away from the real $k$-axis, so that it runs instead along the diagonal $k^\prime = \frac{1-i}{\sqrt{2}}\kappa$ for real $\kappa\in(-\infty,\infty)$. See Fig.~\ref{ContourFig}. Along the long diagonal part of the new trajectory, the large value of $\frac{\hbar\tau}{m\lambda^{2}}$ means that the integrand is tightly concentrated near the origin (where no singularity takes place because $f_l(k^\prime)\rightarrow -a_l k^{\prime 2l}$ near zero \cite{Tayl72}) within $|k^\prime|\lesssim \tau^{-1/2}$, so that the whole contribution of this diagonal trajectory decays with increasing $\tau$ as $\sim \tau^{-1/2}$. This is the transient component of $T_R(\hbar\mathbf{k},\tau)$, which does not concern us. 

The secular component in $T_R(\hbar\mathbf{k},\tau)$ thus comes entirely from the singularity at $k^\prime=k-i\epsilon$, a contribution that we can calculate using Cauchy's Differentiation Formula. Ignoring the transient component given by the diagonal we find\begin{widetext}
\begin{eqnarray}\label{R3}
T_R(\hbar \mathbf{k},\tau) &\equiv &-\frac{4\pi \hbar\tau}{m}\tilde{f}(k,0)  = -\frac{4\pi \hbar\tau}{m}\left( \mbox{Re}[\tilde f(k)] + i\, k\, \frac{\sigma(k)}{4\pi} \right)  \;,
\end{eqnarray}

where 
\begin{eqnarray}
\tilde{f}(k,\theta)&\equiv & \sum_{l=0}^\infty (2l+1)P_{l}(\cos\theta)[1\pm(-1)^{l}]f_{l}(k)
\end{eqnarray}
$\tilde{f}(k,\theta)$ is the bosonic or fermionic scattering amplitude at wave number $k$ for the polar deflection angle $\theta$ and we have used the Optical Theorem to relate its imaginary part with the total scattering cross section $\sigma$. 

For bosons at low momentum we can approximate $\mbox{Re}[\tilde f(q)]$ with the s-wave scattering length

\begin{equation}\label{eq:DefScatteringLength}
\lim_{k\rightarrow 0} f(k,0) = -a
\end{equation}

If we use this approximation for $T_R$ we find a contribution to $\rho_{I}(\mathbf{r},\mathbf{r}',\tau,t_{0})$ of the form

\begin{eqnarray}\label{eq:GrossPitaevskiDensity}
i(N-1)\,(R(\mathbf{r},\mathbf{r}^\prime,\tau,t_{0})-\,R^*(\mathbf{r}^\prime,\mathbf{r},\tau,t_{0})) &= & \frac{4\pi a \hbar \tau}{m}\, i\,\left(n(\mathbf{r})-n(\mathbf{r}^\prime) \right)\, \rho_{I}(\mathbf{r},\mathbf{r}^\prime,0,t_0)+(...)
\end{eqnarray}

where we have assumed for the total particle density in a point  $n(\mathbf{r})= N\, \int d^3 p\, W(\mathbf{r},\mathbf{p})$ that $N-1 \approx N$.\\

In a classical picture, this would correspond to the Vlasov mean-field correction to the Boltzmann Equation \cite{Cercignani75}. In the quantum picture, this is exactly the same term found in the Gross-Pitaevskii Equation for a Bose-Einstein Condensate \cite{Pethick08}. It is simply a nonlinear mean field correction to the Hamiltonian evolution and we will suppose this to be the role of $\mbox{Re}[\tilde f(q)]$ for this term at arbitrary momenta as well.\\

On the other hand, the imaginary part of the scattering amplitude will certainly yield a non-Hamiltonian term

\begin{eqnarray}\label{eq:LossTermInDensityRepresentation}
i\,(R(\mathbf{r},\mathbf{r}^\prime,\tau,t_{0})-\,R^*(\mathbf{r}^\prime,\mathbf{r},\tau,t_{0})) &=& -\frac{4\pi \tau}{\hbar^3} \int d^3p\,e^{-\frac{i}{\hbar}\mathbf{p}\cdot (\mathbf{r}-\mathbf{r}^\prime)}\, W\left(\frac{\mathbf{r}+\mathbf{r}^\prime}{2},\mathbf{p},t_0\right)\, \left(\gamma(\mathbf{r},\mathbf{p},t_0) +  \gamma(\mathbf{r}^\prime,\mathbf{p},t_0) \right)+\,i\,(...)\\
\gamma(\mathbf{r},\mathbf{p},t_0) & =& \int d^3k\, \frac{\hbar k}{m} \, \frac{\sigma(k)}{4\pi}\, W(\mathbf{r},\mathbf{p}+\hbar\mathbf{k},t_0)
\end{eqnarray}

This $\gamma$ is the same one from the classical Boltzmann Equation (\ref{eq:CSALossCompact}).

If we put the expression above (\ref{eq:LossTermInDensityRepresentation}) in the Wigner representation, we obtain finally a form exactly like the loss term $\Gamma_{-}$ from the classical Boltzmann equation

\begin{eqnarray}\label{Gam-}
\mathcal{R}(\mathbf{r},\mathbf{p},\tau,t_{0})&\equiv &\int\!\frac{d^{3}z}{(2\pi)^{3}}\,e^{i\mathbf{p}\cdot\mathbf{z}}\, i\left[R\left(\mathbf{r}+\frac{\mathbf{z}}{2},\mathbf{r}-\frac{\mathbf{z}}{2},\tau,t_{0}\right)-R^{*}\left(\mathbf{r}-\frac{\mathbf{z}}{2},\mathbf{r}+\frac{\mathbf{z}}{2},\tau,t_{0}\right)\right]\\
& = &2\frac{\tau}{m}W(\mathbf{r},\mathbf{p},t_{0})\int\!d^{3}p'\,W(\mathbf{r},\mathbf{p}',t_{0})\int\!d^{3}p''\,\delta\left(|\mathbf{p}''|^{2}-\left|\frac{\mathbf{p}-\mathbf{p}'}{2}\right|^{2}\right)\frac{d\sigma}{d\Omega} + \mbox{(...)} \nonumber\\
& = & \tau\, W(\mathbf{r},\mathbf{p},t_0)\, \gamma(\mathbf{r},\mathbf{p},t_0) \equiv \tau \, \Gamma_-(\mathbf{r},\mathbf{p})\nonumber
\end{eqnarray}

\end{widetext}

Here we have omitted the mean field contribution of $\alpha(q)$ and we have droppped again the terms of order $\lambda/L$ in the position integration variables, as we did in (\ref{R1TWScales}) to obtain (\ref{R1TWScalesLocalInPosition}). 

A similar series of calculations and approximations produces an expression for the secular part of the Wigner representation of the kernel $T_S(\hbar\mathbf{k},\mathbf{z},\tau)$ from (\ref{R1TWScalesLocalInPosition})

\begin{eqnarray}\label{Gam+}
T_S(\hbar\mathbf{k},\mathbf{z},\tau) &=& 8\pi \,\tau\, \mbox{sinc}(2kz)\, \frac{\hbar k}{m}\,\frac{d\sigma}{d\Omega}\,\Theta\left(\frac{\hbar k \tau}{m}-2z\right) \nonumber 
\end{eqnarray}

The $\Theta$-function reminds that the gain term is only relevant close to the diagonal within a range determined by the post-scattering reach, i.e. the distance that particles with wave number $q$ can travel after being scattered within the time $\tau$. Although this region of relevance is very narrow this term should not be plainly neglected because it ensures particle conservation. Indeed, by ignoring this on-diagonal restriction, and rewriting this term in the Wigner representation we obtain exactly the Boltzmann equation's gain term $\Gamma_+$. It is written in a more compact form without using Wigner as

\begin{widetext}
\begin{eqnarray}
S(\mathbf{x},\mathbf{x}^\prime) &=& 8\pi\tau \int d^3s \,\xi(\mathbf{s},|\mathbf{x}-\mathbf{x}^\prime|) \,\left| \rho_I\left( \frac{\mathbf{x}+\mathbf{x}^\prime+\mathbf{s}}{2}, \frac{\mathbf{x}+\mathbf{x}^\prime-\mathbf{s}}{2}\right) \right|^2 \nonumber\\
\xi(\mathbf{x},z) & = & \int d^3q\, \frac{q}{m}\, \mbox{sinc}(2q z)\,e^{-2i\mathbf{q}\cdot \mathbf{s}}\, \frac{d\sigma(\mathbf{q})}{d\Omega}\Theta(q\tau/m-z)
\end{eqnarray}
\end{widetext}

If we then differentiate (\ref{R1T}) with respect to $\tau$ after discarding its transient and mean field terms, we obtain the classical Boltzmann equation precisely for the quantum Wigner function. Putting all of the terms together

\begin{widetext}
\begin{eqnarray}\label{eq:QBEfullClassicalGas}
\frac{d\rho}{dt} &=& -\frac{i}{\hbar}\langle \mathbf{x}| [\hat H_0,\hat \rho_1]| \mathbf{x}^\prime\rangle\nonumber\\
 & -& 4\pi i \int d^3p\, \left[\mu(\mathbf{x},\mathbf{p})-\mu(\mathbf{x}^\prime,\mathbf{p})\right]\, e^{i\mathbf{p}\cdot (\mathbf{x}-\mathbf{x}^\prime)}\, W\left( \frac{\mathbf{x}+\mathbf{x}^\prime}{2},\mathbf{p}\right)\nonumber\\
 & - & 4\pi\,  \int d^3p\,\left[\gamma(\mathbf{x},\mathbf{p})+\gamma(\mathbf{x}^\prime,\mathbf{p})\right]\, e^{i\mathbf{p}\cdot (\mathbf{x}-\mathbf{x}^\prime)}\, W\left( \frac{\mathbf{x}+\mathbf{x}^\prime}{2},\mathbf{p}\right)\nonumber\\
 & +& 8\pi \int d^3s \,\xi(\mathbf{s},|\mathbf{x}-\mathbf{x}^\prime|) \,\left| \rho_I\left( \frac{\mathbf{x}+\mathbf{x}^\prime}{2}+\frac{\mathbf{s}}{2}, \frac{\mathbf{x}+\mathbf{x}^\prime}{2}-\frac{\mathbf{s}}{2}\right) \right|^2
\end{eqnarray}

The following auxiliary functions were defined along the paper for a shorter notation

\begin{eqnarray}
\mu(\mathbf{x},\mathbf{p})&=& \int d^3q\, \mbox{Re}[\tilde f(q)]\, W(\mathbf{x},\mathbf{p}+2\mathbf{q})\\
\gamma(\mathbf{x},\mathbf{p}) &=& \int\!d^3q\,\frac{q}{m} \frac{\sigma(q)}{4\pi}\,  W(\mathbf{x},\mathbf{p}+2\mathbf{q})\\
\xi(\mathbf{s},z) &=& \int d^3q\, \frac{q}{m}\, \mbox{sinc}(2q z)\,e^{-2i\mathbf{q}\cdot \mathbf{s}}\, \frac{d\sigma(\mathbf{q})}{d\Omega}\Theta(q\tau/m-z)
\end{eqnarray}
\end{widetext}

The first line in (\ref{eq:QBEfullClassicalGas}) describes the particle's flight under the noninteracting Hamiltonian and it is equivalent to the Moyal expansion; the second one is the nonlinear correction to the Hamiltonian that yields Gross-Pitaevskii for cold bosons; and the last two correspond to the loss and gain terms in the CBE.

With this we have shown that both the Gross-Pitaevskii Equation and the classical Boltzmann Equation can be recovered from our time-dependent scattering model. But more interesting, we can apply our analysis to situations far from the semiclassical situation considered until now.

\section{Beyond Boltzmann: Decoherence}\label{section:Decoherence}

So far we have studied a dilute gas fulfilling certain restrictions upon its length scales that guarantee the validity of the {\it local and instantaneous} scattering approach. These conditions are certainly matched by a semiclassical gas and we have consequently shown that in such cases the equation of motion is the CBE plus some additional mean field corrections.
But let us deviate now towards a regime that is more clearly identified with quantum behavior. A very simple - and legitimately quantum - scenario for a single particle would be having it trapped inside a large potential well, with walls strong and thick enough to avoid the particle to trespass them, and at some point lowering and shortening one of the walls only, in such a way that the wall is still strong enough to forbid the particle to cross it by thermal hopping (i.e. the kinetic energy of the particle is still lower than the potential wall height) but it is thin enough so that the particle has a non-neglectable probability to tunnel through it. In this particular scenario any trapped particle that hits the weakened wall will split into a Schr\"odinger's Cat state, i.e. a superposition between being reflected by the wall and scaping to the outside by tunneling. This states would have the form

\begin{eqnarray}\label{def:TunnelingCatStates}
\phi_{k}(x) &=& \Theta(-x)\,\left(e^{ikx}+r_k \, e^{-ikx} \right)+\Theta(+x)\,t_k\, e^{ikx}
\end{eqnarray}

\begin{figure}
\includegraphics[width=0.4\textwidth]{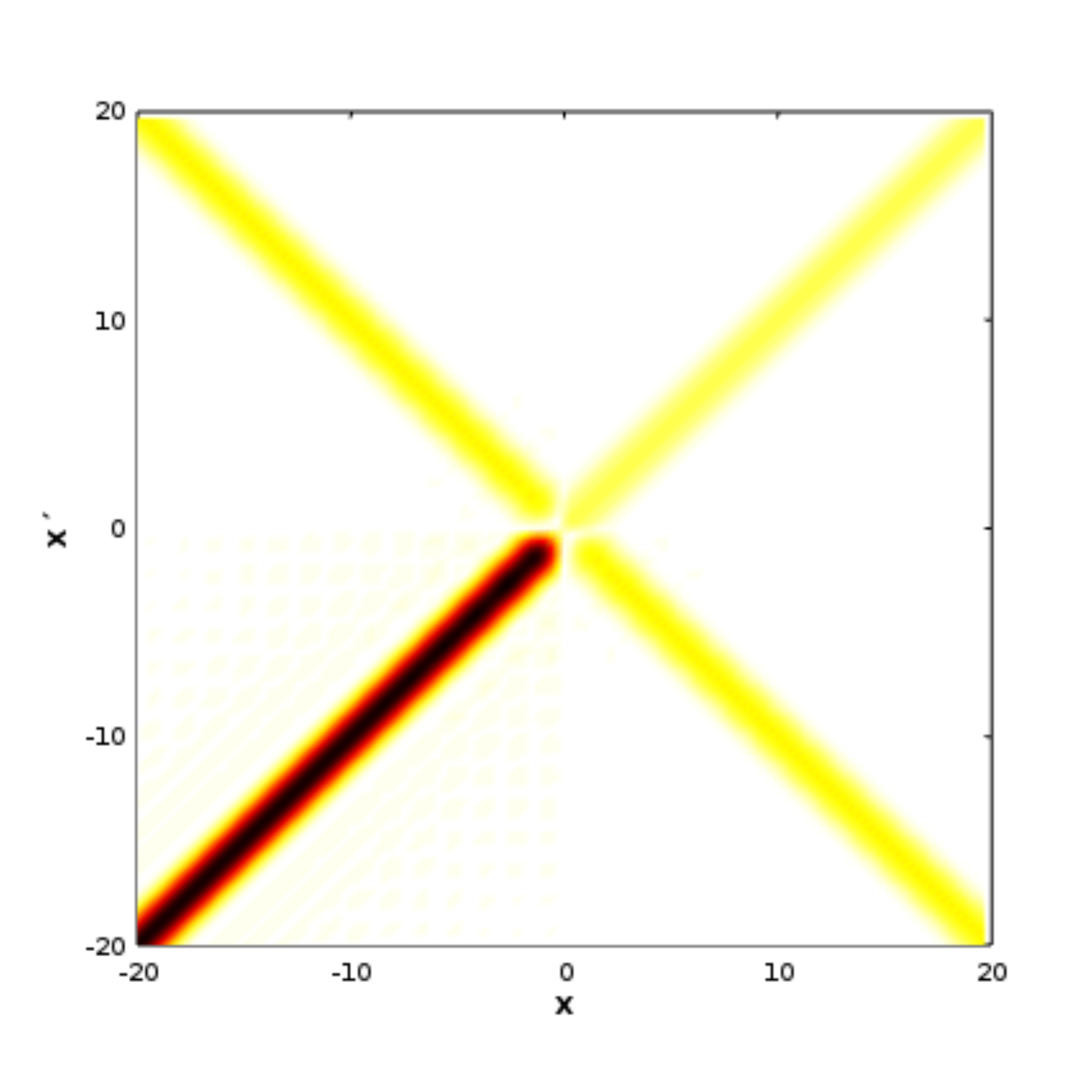}
\caption{Example of one body density matrix (absolute value $|\rho(\mathbf{x},\mathbf{x}^\prime)|$ represented) corresponding to an ideal gas initially in a Boltzmann distribution that tunnels through a potential barrier of neglectable thickness at $x=0$ (see (\ref{def:TunnelingCatStates}) and (\ref{def:BoltzmannDistribution})). The off-diagonal $x^\prime = -x$ represents the non-vanishing long range coherence between opposite sides of the weakened wall and corresponds to the term with $X(\mathbf{r},\mathbf{q})$ from (\ref{eq:AnsatzLongRange}).}
\label{fig:LongRangeCoherence}
\end{figure}

Now that we imagined this situation for one particle, imagine it for a gas of noninteracting particles: each one of the particles can hit the wall and split into an entangled state, a superposition between remaining inside the container and scaping it. An example of the single particle density matrix that can be obtained in this situation is depicted in Fig.\ref{fig:LongRangeCoherence}, with a Boltzmann distribution

\begin{eqnarray}\label{def:BoltzmannDistribution}
n_k & \propto & e^{-k^2/2\lambda^2}
\end{eqnarray}

of Schr\"odinger's Cat states $\phi_k$, so that the reduced density matrix in the steady state of this particular problem would be

\begin{eqnarray}\label{def:RDMTunnelingThrottling}
\rho^0(\mathbf{x}|\mathbf{x}^\prime) &=& \int dk\, n_k\, \phi_k^*(\mathbf{x})\, \phi_k(\mathbf{x}^\prime)
\end{eqnarray}

Its main feature is that it shows a very long range coherence that will not vanish. The gas has no external bath to couple and, as we said, particles in the gas do not interact with each other, keeping this long range coherence unaltered at arbitrary separations from the barrier.

Suppose now that, while keeping the gas isolated from any external influence, we do turn on the interactions between the gas' particles. Will this long-range coherence survive? Applying our two body collision analysis we will show here that the answer is `no'. With this analysis a new route to quantum decoherence is open, an alternative to the conventional view of quantum decoherence as a consequence of coupling to external sources.

To express mathematically the kind of long-range decoherence that we are explaining with this model, let us consider now the case of a correlation function that, besides a component close to local equilibrium described by $W$ in the previous section, includes a long-range coherence term $X$ such that

\begin{eqnarray}\label{eq:AnsatzLongRange}
\rho_I(\mathbf{r},\mathbf{r}^\prime, t)& =& \int d^3 p\,W\left(\frac{\mathbf{r}+\mathbf{r}^\prime}{2}, \mathbf{p},t\right)\, e^{\frac{i}{\hbar} \mathbf{p} \cdot (\mathbf{r}-\mathbf{r}^\prime)} \nonumber\\
& +& \int d^3 p\,X\left(\frac{\mathbf{r}-\mathbf{r}^\prime}{2},\mathbf{p},t\right)\, e^{\frac{i}{\hbar} \mathbf{p} \cdot (\mathbf{r}+\mathbf{r}^\prime)}
\end{eqnarray}
and again both contributions $W$ and $X$ will be taken to be smooth in one of their arguments in comparison with the other one
\begin{eqnarray}
X(\mathbf{r},\mathbf{p},t) &=& \tilde{X}\left(\frac{\mathbf{r}}{L},\frac{\lambda \mathbf{p}}{\hbar},t\right)\nonumber\\
W(\mathbf{r},\mathbf{p},t) &=& \tilde{W}\left(\frac{\mathbf{r}}{L},\frac{\lambda\mathbf{p}}{\hbar},t\right)\nonumber
\end{eqnarray}

The expression above would be a generalization that covers the example given in (\ref{def:RDMTunnelingThrottling}). Since the long-range coherence is expected to vary little along the off-diagonal but fast in the direction paralel to the diagonal, we have chosen to represent it with an atypical {\it anti}-Wigner transform: if the Wigner function is a Fourier transform along the off-diagonal, this alternative transform is Fourier along the diagonal.\\

Our next goal is to determine how is the time evolution of a correlation function of this form influenced by collisions, focusing specially on the evolution of the long range coherence part $X$. Since the equation of motion is quadratic in $\rho_I$, we can expect three kind of terms to appear. First, quadratic terms in $X$ can be discarded if the magnitude of $X$ is smaller than that of $W$: we will only consider linear perturbations in $X$ here. Second, quadratic terms in $W$ have been already taken into account previously and we will not need to repeat those calculations. Finally, it is the crossed terms linear in both $W$ and $X$ that are going to be of our interest, since they are neither taken into account already nor small enough to be neglected when it comes to the time evolution of $X$.

Once we have only kept linear terms in $X$, there are again two kind of contributions to take into account, the $R$-terms and the $S$-terms, as in Equation (\ref{R1T}). But the $S$-terms are irrelevant for the long range coherence because they only matter around the diagonal for a very short distance, i.e. the post-scattering reach. Then we can discard them as well.

\begin{widetext}
We are left only with $R$-terms that are linear in $X$. There are two contributions of this form

\begin{eqnarray}
R(\mathbf{r},\mathbf{r}^\prime) &=& \int d^3p d^3q d^6s_j\, W\left(\frac{\mathbf{r}+\mathbf{r}^\prime}{2}-\frac{\mathbf{s}_1-\mathbf{s}_2}{4},\mathbf{p}\right)\, X\left(\frac{\mathbf{s}_1-\mathbf{s}_2}{2},\mathbf{q}\right)\, e^{\frac{i}{\hbar}\mathbf{p}\cdot (\mathbf{r}-\mathbf{r}^\prime)}\, e^{2\frac{i}{\hbar}\mathbf{q}\cdot \mathbf{r}} \nonumber \\
& & \qquad \times \, e^{\frac{i}{2\hbar} \mathbf{s}_1 \cdot (\mathbf{p}+3\mathbf{q})}\, e^{\frac{i}{2\hbar}\mathbf{s}_2\cdot (\mathbf{p}-\mathbf{q})}\, [T(\mathbf{s}_1,\mathbf{s}_2)\pm T(\mathbf{s}_1,-\mathbf{s}_2)]\nonumber\\
& & + \int d^3p d^3q d^6s_j\, W\left(\mathbf{r}-\frac{3\mathbf{s}_1+\mathbf{s}_2}{2},\mathbf{p}\right)\, X\left(\frac{\mathbf{r}-\mathbf{r}^\prime}{2}-\frac{\mathbf{s}_1-\mathbf{s}_2}{4},\mathbf{q}\right)\, e^{\frac{i}{\hbar}\mathbf{q}\cdot (\mathbf{r}+\mathbf{r}^\prime)}\nonumber \\
& &\qquad \times \, e^{\frac{i}{2\hbar} \mathbf{s}_1 \cdot (\mathbf{p}-\mathbf{q})}\,e^{-\frac{i}{2\hbar}\mathbf{s}_2\cdot (\mathbf{p}-\mathbf{q})}\,[T(\mathbf{s}_1,\mathbf{s}_2)\pm T(\mathbf{s}_1,-\mathbf{s}_2)]
\end{eqnarray}

For simplicity time arguments were ommited in this expression. Now we apply the same argument about locality used before to ignore dummy integral variables in the first argument of $W$ and $X$.

\begin{eqnarray}
R(\mathbf{r},\mathbf{r}^\prime) &=& \int d^3p d^3q \, W\left(\frac{\mathbf{r}+\mathbf{r}^\prime}{2},\mathbf{p}\right)\, X\left(0,\mathbf{q}\right)\, e^{\frac{i}{\hbar}\mathbf{p}\cdot (\mathbf{r}-\mathbf{r}^\prime)}\, e^{2\frac{i}{\hbar}\mathbf{q}\cdot \mathbf{r}}\, \int d^6s_j\, e^{\frac{i}{2\hbar} \mathbf{s}_1 \cdot (\mathbf{p}+3\mathbf{q})}\,e^{\frac{i}{2\hbar}\mathbf{s}_2\cdot (\mathbf{p}-\mathbf{q})}\, [T(\mathbf{s}_1,\mathbf{s}_2)\pm T(\mathbf{s}_1,-\mathbf{s}_2)]\nonumber\\
& & + \int d^3p d^3q \, W\left(\mathbf{r},\mathbf{p}\right)\, X\left(\frac{\mathbf{r}-\mathbf{r}^\prime}{2},\mathbf{q}\right)\, e^{\frac{i}{\hbar}\mathbf{q}\cdot (\mathbf{r}+\mathbf{r}^\prime)}\, \int d^6s_j\, e^{\frac{i}{2\hbar} \mathbf{s}_1 \cdot (\mathbf{p}-\mathbf{q})}\,e^{-\frac{i}{2\hbar}\mathbf{s}_2\cdot (\mathbf{p}-\mathbf{q})}\, [T(\mathbf{s}_1,\mathbf{s}_2)\pm T(\mathbf{s}_1,-\mathbf{s}_2)]\nonumber
\end{eqnarray}

We can again identify kernels of the form of $T_j$ as in Eq. (\ref{R1TWScalesLocalInPosition}). In the first line, the kernel is the overlapping between a plane wave of momentum $3\mathbf{p}-\mathbf{q}$ and the outgoing scattered wave of momentum $\mathbf{p}-\mathbf{q}$: this yields no secular contribution. In turn, the second line correspond to the case of equal momenta, one that is already known as the kernel $T_R$. This gives a final contribution of the form

\begin{eqnarray}\label{eq:LinearLongRange}
R(\mathbf{r},\mathbf{r}^\prime,\tau,t_0) &=& \tau\int d^3p\, X\left(\frac{\mathbf{r}-\mathbf{r}^\prime}{2},\mathbf{p},t_0\right)\, e^{\frac{i}{\hbar}\mathbf{p}\cdot (\mathbf{r}+\mathbf{r}^\prime)}  \int d^3q\, W(\mathbf{r},\mathbf{p}+2\mathbf{q},t_0)\, T_R(\mathbf{q},\tau)
\end{eqnarray}

\end{widetext}

We can apply the same kind of {\it anti}-Wigner transformation that we used to define $X$ in (\ref{eq:AnsatzLongRange}) together with the locality argument to obtain an approximated linear equation of motion for $X$ alone

\begin{eqnarray}\label{eq:LongRangeDecoBoltzmann}
\dot X(\mathbf{r},\mathbf{p}) &\approx & - 4\pi(N-1)\, \sum_\pm \gamma (\pm\mathbf{r},\mathbf{p}) \, X(\mathbf{r},\mathbf{p})
\end{eqnarray}

where again the same dependence in the $\gamma (\mathbf{r},\mathbf{p})$ factor appears. Since $\gamma$ has a clear meaning in the classical context, this suggests a simple and intuitive interpretation: in the limit where collisions can be considered local and instantaneous, decoherence between two points $\mathbf{r}$ and $\mathbf{r}^\prime$ of a gas cloud is proportional to the combined rates of collisions in each one of those points and independent of the distance separating them.

As long as the gas leaking out through the potential barrier provides a steady flux of incoming particles we should not expect the coherence terms in the region $x>0$ and $x^\prime<0$ ($x<0$ and $x^\prime>0$) to disappear completely. A particle that approaches the barrier and splits in a Schr\"odinger's Cat state would not decohere instantly and, until it loses that coherence, the tunneled and the reflected parts of this state will fly away from the barrier during that time. This is, coherence generated in a single tunneling attempt would survive a distance $x_k = \frac{\hbar k}{m} \, \frac{1}{\gamma_k}$ and the coherence regions would display a decaying off-diagonal tail of the form

\begin{eqnarray}
\rho(x|x^\prime) &\propto & \Theta(x)\Theta(-x^\prime) \, e^{-\frac{\gamma_k m}{\hbar k}|x-x^\prime |}
\end{eqnarray}

instead of the infinite range coherence terms shown in Fig.\ref{fig:LongRangeCoherence} for the noninteracting scenario.

Finally, it is worth to see that removing completely coherence terms in the steady flux state of our tunneling toy model transforms the reduced density matrix $\rho$ from the initial Boltzmann distribution with eigenvalues $n_k$ from (\ref{def:BoltzmannDistribution}) and eigenstates $\phi_k$ (\ref{def:TunnelingCatStates}) that are defined for both regions - left {\it and} right of the barrier - into another reduced density matrix with two sets of eigenstates that are only defined in one of the regions - left {\it or} right - with modified eigenvalues

\begin{eqnarray}\label{eq:ModifiedEigenvaluesAfterDecoherence}
n_k \rightarrow \, n_k |t_k|^2/2\, \mbox{(left), or }\, n_k (1+|r_k|^2)/2\,  \mbox{(right)}
\end{eqnarray}

The factors of $1/2$ are the result of the new normalization of the eigenvectors. Being able to determine the eigenvalues of both the coherent and incoherent counterpart allows us to calculate the entropy increase

\begin{eqnarray}\label{eq:EntropyProduction}
\Delta S_k &=& -\frac{|t_k|^2}{2} \log{\left(\frac{|t_k|^2}{2}\right)} \nonumber\\
& & - \left(1-\frac{|t_k|^2}{2}\right) \log{\left(1-\frac{|t_k|^2}{2}\right)}
\end{eqnarray}

per particle with momemtum $k$. This coincides with the thermodynamical entropy increase of a free expansion in two boxes of equal volume for $|t_k|^2=1$.


\section{Conclusions}

In the present paper we have tried to rebuild from quantum mechanical first principles a foundational result of Statistical Mechanics, the Boltzmann Equation. And this has lead immediately to a quite provocative result about quantum decoherence.

Our rederivation of the Boltzmann Equation under quantum formalism looks back to the overseen distinction between different time scales involved in the process of quantum scattering of particles in a gas. We have shown how critical it is to remember that the evolution time $\tau$ considered here must be taken sometimes as zero and sometimes as infinity. This statement, that may sound as trickery, is nothing but a well known principle of multiple-scale analysis. Or in more familiar terms to even the newest students: how important it is sometimes to take double limits in the right order.
Had we taken an infinite time limit from the beginning (this is, had we applied the usual scattering operators), then we would have found an infinite post-scattering reach. Indeed, this has been done to find the so called Linear Boltzmann Equation \cite{Horn06} - and it is not wrong to do so because the post-scattering position of the bath particles is meaningless in that context. In turn, this would be a dire mistake in our case because there is no clear distinction between bath and test particles, they are all part of the same gas of indistinguishable particles. Infinite post-scattering reach would have derived into unrealistic `teleportation' of particles because the applied scattering operator would have included an infinite time of flight.
On the other hand, if we had taken the time limit to zero from the beginning then we would not have been able to distinguish the secular behavior. The only effect found in that limit is a mean field correction to the Hamiltonian and decoherence would have been overseen.
We notice too that this time hierarchy implies a length hierarchy: interaction range, post-scattering reach and typical interparticle, must be, in that precise order, one larger than another. This last fact stablishes a very well defined range of applicability of this formalism. It means that we are bound to dilute gases with short range interactions. 

The dilute gas approximation has been indeed one of our most often wielded arguments here. More concretely, it is by arguing that only two particles will ever collide at once that we can use the approximation (\ref{dilute}) to discard higher order Green functions $L_n$ in the expansion (\ref{Ansatz}). But even more important, we use (\ref{SZA}) as a closure relation to infer from $\rho_1$ the information about $\rho_2$ that we cannot really access. The latter one is our strongest assumption, it implies to acknowledge that information from the system is leaked towards non-accessible higher order correlations. This loss of information is assumed rather than justified and regarding our results concerning quantum decoherence of $\rho_1$ we can say that such higher order correlations play the role of the `bath' used in most works about decoherence found in the literature \cite{Zurek03}.

We must nevertheless warn the readers about the extreme simplicity of (\ref{SZA}). This is precisely why we have used it here: it resembles the Classical Stosszahl Ansatz, only patched to fit bosons or fermions. But that classical version was meant for probability distributions, and, as we sketched in our discussion about the Wigner function, classical probability distributions do not have an exact counterpart in the quantum formalism. We suspect that a careful quantum treatment for dilute but degenerate quantum gases should not use this ansatz but a more elaborate one instead. Here the only implemented difference between bosons and fermions concerns their scattering amplitudes but as soon as we switch to a better closure relation or include spin in the gas' model more differences should appear and our simplification to $\tilde f$ will not suffice. This will be specialy true for degenerate fermions, although quite on the contrary, for bosons, it has been almost trivial to derive a density matrix version of the Gross-Pitaevskii Equation for Bose-Einstein Condensates (\ref{eq:GrossPitaevskiDensity}) \cite{Pethick08}. It is easy to see now, with the benefit of hindsight, that this is because (\ref{SZA}) fits a condensate very accurately.

All the previous considerations are important and should be regarded if this method is to be applied to further cases. But as soon as we have tried to test our formalism in a genuine quantum situation a much more provocative result have been found: a new view and mechanism of decoherence for macroscopic systems.

An important aspect to consider is the dependence with position in contrast to the dependence with separation. In some of the first works about decoherence \cite{CaldeiraLeggett83} \cite{JoosZeh85} the system was subject to coupling with an homogeneous bath and a direct quadratic dependence with the separation distance $|\mathbf{x}-\mathbf{x}^\prime|$ was found. This growing dependence with distance was later discovered to saturate \cite{GallisFleming90} \cite{AnglinPazZurek97} \cite{Gilz14}. We find instead a decay factor $\gamma(\mathbf{r},\mathbf{p})$ that is position dependent and insensitive to the distance. This is easily explained by two facts. First, we differentiate between the collision event, i.e. while the particle flight is still influenced by the interaction, and the post-scattering flight (and its associated distance, the post-scattering reach). We take the later one to be much longer than the first one that consequently take the first one to be zero and that leave us way into the distance independent regime but we suspect that a more careful treatment of the gain term (the last line in (\ref{eq:QBEfullClassicalGas})) may well recover this short range dependence. And second, contrary to our predecessors we are considering a bath that is not necessarily homogeneous. Our bath is the gas itself and it is not necessarily neither homogeneous nor even in equilibrium. But this second difference points out towards a very intuitive principle about how coherence is destroyed: while quantum coherence can be {\it global}, i.e. with an arbitrary long range, its suppression may hapen {\it locally}. Given a long range Schr\"odinger's cat state with two sides, each one of its sides will suffer separately from collisions with other particles, being the total loss of coherence the combination of both, as we read in the third line of (\ref{eq:QBEfullClassicalGas}). Although in our case these collisions are with the gas' particles themselves, it is quite reasonable to think that this is a more general principle: that coherences generated locally can be destroyed globally.

Finally, but not less important, we must emphasize that the mechanism of decoherence here described is based in a conception of quantum coherence that may not be familiar to all readers, i.e. intrinsic or self induced decoherence. Most previous works consider decoherence only in the context of coupling the system to an external bath. This can project the false image of decoherence to be a disease carried by classical systems that infects quantum ones turning them into carriers as well. Instead of this, to generalize decoherence for isolated systems, we only care about its coarse-grained representation, which in our case is the one body reduced density matrix, and we consequently claim that coherence is lost when this representation loses it. It does not matter whether the gas started in a pure state of the N-body Hilbert space: although such a state would remain `pure' because the time evolution is given by the N-body Schr\"odinger equation, the coarse-grained representation of it will not. Stating that the gas remains in a pure state because it is isolated is just as useful as its classical counterpart. This is, like saying that the time evolution of a gas of classical particles is reversible because the microscopic dynamics of the particles is. It is a well stablished fact that, in the coarse-grained description taken in Statistical Mechanics and Thermodynamics, its evolution is not reversible and this same view can be taken when talking about quantum macroscopic systems if we accept that macroscopic can also mean ``composed by a macroscopic number of parts''. 

Concerning the possibility to test these results, we can first look back to previous experiments. Consider for instance \cite{Ott04}, where bosonic/fermionic atoms where loaded in a vertical lattice and released to look for interference patterns. It was observed that the posibility of bosons to collide, in opposition to polarized fermions, destroyed rapidly the coherence necessary to generate interference patterns. Back then it was attributed to dephasing --- but the  collisional decoherence among gas particles described in this work would have a similar effect, also differentiating bosons, with low momentum s-wave scattering as its main contribution, and fermions, with suppressed s-wave scattering.

About the concrete system we have picked up in Section \ref{section:Decoherence} to illustrate the decoherence mechanism, it is also possible to realize experimentally. A sample of cold gas could be loaded in an optical trap with the shape of a large double well, almost flat, populating only one of the sides and making the potential wall between them strong enough to avoid both thermal and tunneling leaking. Then lower that barrier to let the gas leak for a while, strengthen the barrier again and let both sides to ``cooldown''. After that, remove the barrier and push both samples toghether again and check interference patterns to test how coherent they remained. This could be repeated adjusting interaction strengths between particles using  Feshbach resonances and varying the cooldown times or the barrier. Inhomogeneus traps may give a chance to explore the role of local density.
The system proposed could be interpreted as a special case of quantum free expansion problem, except that instead of removing the wall completely we have a quantum analog of a porous layer. Two possible outcomes can be easily expected. Either decoherence suppresses tunneling altoghether, avoiding the gas to scape far away from the barrier or else tunneled particles can scape, a free expansion process takes place and the suppressed coherence can be interpreted as (thermodynamical) entropy generation, as is done for example in \cite{Kim2017}. Consequently, this would be a case where the line distinguishing quantum decoherence and thermodynamic equilibration disappears. In this precise system they are the same process.



\begin{appendix}

\begin{widetext}
\newpage


\section{Reduction of $\tilde{L}_{2}$ to the single-particle finite-time $T$-matrix}\label{Appendix:FromL2toT}
While it is clear from our expansion of $\hat{a}$ in (\ref{Ansatz}) that $\tilde{L}_2$ must somehow represent the effects of collisions between two particles, the precise way that $\tilde{L}_2$ does this involves the question of how one would have to re-arrange particles at time $t_0$ in order to make the removal of a particle at $t_0$ produce the same effect as removing the particle at a later time $t_0+\tau$. Even the most formally minded theorist should admit that this is a weird question. Particles are conserved in non-relativistic interactions, and never actually removed. The formalism of second quantization is nonetheless a powerful one and the answers to its unphysical questions can be useful. Statements about what operations at earlier times would produce the same effects at later times as removing a particle from position $\mathbf{r}$ turn out to be efficient ways to describe gas dynamics.

Inserting (\ref{Ansatz}) into the Heisenberg equation of motion (\ref{Heis1}), and applying the canonical (anti)commutation relation (\ref{CCR}), reveals this \emph{exact} equation for $L_{2}$ (without the tilde accent!):
\begin{eqnarray}
\left[i\hbar\frac{\partial}{\partial t}+\frac{\hbar^{2}}{2m}\nabla_{r}^{2}\right]  L_{2}(\mathbf{r};\mathbf{q}';\mathbf{q}_{1},\mathbf{q}_{2};t) 
&=& \int\!d^{3}z\,d^{3}z'\,U(\mathbf{z}-\mathbf{r})\,G_{1}^{*}(\mathbf{z},\mathbf{q}',t)\,G_{1}(\mathbf{z},\mathbf{z}',t)\,L_{2}(\mathbf{r};\mathbf{z}';\mathbf{q}_{1},\mathbf{q}_{2};t)\nonumber\\
&&+\int\!d^{3}z\,U(\mathbf{z}-\mathbf{r})\,G_{1}^{*}(\mathbf{z},\mathbf{q}',t)\,G_{1}(\mathbf{z},\mathbf{q}_{2},t)\,G_{1}(\mathbf{r},\mathbf{q}_{1},t)\;.\label{L2eq}
\end{eqnarray}
Demanding also that $\hat{a}(\mathbf{r},t_{0})$ is equal to itself, we find the initial condition
\begin{eqnarray}\label{initconds}
L_{2}(\mathbf{r};\mathbf{q}';\mathbf{q}_{1},\mathbf{q}_{2};0) &=& 0\label{initconds2}\;.
\end{eqnarray}

We can then express $L_{2}$ exactly in terms of another function, defined as
\begin{eqnarray}\label{propagator2body}
	G_{2}(\mathbf{r},\mathbf{r}';\mathbf{q}_{1},\mathbf{q}_{2};t) &=& G_{1}(\mathbf{r},\mathbf{q}_{1},t)G_{1}(\mathbf{r}',\mathbf{q}_{2},t) \\
 && \!\!\!\!\! \!\!\!\!\!+ \int\!d^{3}q'\,G_{1}(\mathbf{r}',\mathbf{q}',t)L_{2}(\mathbf{r};\mathbf{q}';\mathbf{q}_{1},\mathbf{q}_{2};t)\nonumber
\end{eqnarray}
Together with (\ref{L2eq}), this implies that $G_{2}$ obeys
\begin{eqnarray}\label{G2eq}
\left[i\hbar\frac{\partial}{\partial t}+\frac{\hbar^{2}}{2m}(\nabla^{2}_{r}+\nabla^{2}_{r'})+U(\mathbf{r}-\mathbf{r}^{\prime})\right]G_2 = 0\;,
\end{eqnarray}
while from (\ref{initconds2}) we find the initial condition
\begin{eqnarray}\label{G2init}
	G_{2}(\mathbf{r},\mathbf{r}^{\prime};\mathbf{q}_{1},\mathbf{q}_{2};0) = \delta^{3}(\mathbf{q}_{1}-\mathbf{r})\,\delta^{3}(\mathbf{q}_{2}-\mathbf{r}^{\prime})\;.
\end{eqnarray}
Equations (\ref{G2eq}) and (\ref{G2init}) mean that $G_{2}$ is precisely the propagator for the two-particle Schr\"odinger initial value problem: for any initial 2-particle wavefunction $\psi_{I}(\mathbf{r},\mathbf{r}^{\prime})$, the time-dependent wave function
\begin{eqnarray}\label{2particle}
	\psi(\mathbf{r},\mathbf{r}^{\prime},t) = \int\!d^{6}q_{j}\,G_{2}(\mathbf{r},\mathbf{r}^{\prime};\mathbf{q}_{1},\mathbf{q}_{2};t-t_{0})\,\psi_{I}(\mathbf{q}_{1},\mathbf{q}_{2})\nonumber
\end{eqnarray}
satisfies
\begin{eqnarray}
	i\hbar\frac{\partial}{\partial t}\psi&=&\left[-\frac{\hbar^{2}}{2m}(\nabla^{2}_{r}+\nabla^{2}_{r'})+U(|\mathbf{r}-\mathbf{r}^{\prime}|)\right]\psi\; \nonumber\\
\psi(\mathbf{r},\mathbf{r}^{\prime},t_{0})&=&\psi_{I}(\mathbf{r},\mathbf{r}^{\prime})\;.\nonumber
\end{eqnarray}

$G_{2}$ can therefore be found as usual by converting to center-of-mass and relative co-ordinates, to reduce exactly to an effective one-particle problem for a particle with the reduced mass $m/2$. We write
\begin{eqnarray}\label{G2g}
G_{2}(\mathbf{r},\mathbf{r}^{\prime};\mathbf{q}_{1},\mathbf{q}_{2};t) &=& \left(\frac{m}{\pi i \hbar t}\right)^{\frac{3}{2}}e^{\frac{i m}{\hbar t}\left|\frac{(\mathbf{q}_{1}+\mathbf{q}_{2})-(\mathbf{r}+\mathbf{r}^{\prime})}{2}\right|^{2}}\,\times\, g(\mathbf{r}-\mathbf{r}^{\prime},\mathbf{q}_{1}-\mathbf{q}_{2};t)
\end{eqnarray}
and find that we satisfy (\ref{G2eq}) and (\ref{G2init}) if $g$ satisfies the single-particle time-dependent Schr\"odinger equation
\begin{eqnarray}\label{geq}
	i\hbar\frac{\partial}{\partial t}g(\mathbf{s},\mathbf{s}';t) = \left[-\frac{\hbar^{2}}{m}\nabla^{2}_{s}+U(|\mathbf{s}|)\right]g(\mathbf{s},\mathbf{s}',t)
\end{eqnarray}
with the initial condition
\begin{eqnarray}\label{ginit}
	g(\mathbf{s},\mathbf{s}';0) = \delta^{3}(\mathbf{s}-\mathbf{s}')\;.
\end{eqnarray}
In other words, $g(\mathbf{s},\mathbf{s}';t-t_0)$ is the propagator for the single-particle Schr\"odinger equation (\ref{geq}) with arbitrary initial conditions at $t=t_0$. Since this $g$ can be found explicitly, by solving the textbook problem of scattering in a central force, our entire problem is solved in principle---and will be solved explicitly just below.

Before that, we briefly return to $\tilde{L}_2$ and show its relation to the one-body T-matrix. Converting from $L_2$ to the interaction picture $\tilde{L}_2$ by convolving with $G_1$, we find the relation between it and the interaction picture representation of the relative coordinate propagator T
\begin{eqnarray}\label{RelationL2toT}
\tilde{L}_{2}(\mathbf{r},\mathbf{q}';\mathbf{q}_{1},\mathbf{q}_{2};t) &=& i\delta^3\left(\frac{\mathbf{r}+\mathbf{q}'}{2}-\frac{\mathbf{q}_1+\mathbf{q}_2}{2}\right)T(\mathbf{r}-\mathbf{q}',\mathbf{q}_1-\mathbf{q}_2,t)\\
T(\mathbf{s},\mathbf{s}',t)&=&i\delta^3(\mathbf{s}-\mathbf{s}')-i\left(\frac{im}{4\pi\hbar t}\right)^{\frac{3}{2}}\int\!d^3\,s''\,e^{-i\frac{m}{4\hbar t}|\mathbf{s}-\mathbf{s}''|^2}g(\mathbf{s}'',\mathbf{s}',t)\;.
\end{eqnarray}
This is, the kernel $\tilde{L}_2(\mathbf{r},\mathbf{q}';\mathbf{q}_{1},\mathbf{q}_{2};t)$ is a sort of partial propagator from the two initial positions $\mathbf{q}_1,\mathbf{q}_2$ at time $t_0$ to the two final positions $\mathbf{r},\mathbf{q}'$ at time $t_0+t$. Since the center of mass always moves ballistically, without any effect from the interaction force, the elimination of ballistic motion from the interaction picture makes the center of mass remain fixed; hence the delta function in the first line of (\ref{RelationL2toT}). The relative coordinate is propagated non-trivially, by the kernel $T$, which is related to $g$. 

In particular, we can recognize that when the single-particle problem (\ref{geq}) is written in bra-ket notation in the position representation, then
\begin{eqnarray}\label{Tmatrix2}
T(\mathbf{s},\mathbf{s}',t) &=& -i\langle\mathbf{s}|[\hat{U}^\dagger_0(t,0)\hat{U}(t,0)\hat{U}_{0}^\dagger(0,0)-1]|\mathbf{s}'\rangle\nonumber\\
&\equiv& -i\langle\mathbf{s}|[\hat{S}(t,0)-1]|\mathbf{s}'\rangle 
\end{eqnarray}
where $\hat{U}(t_1,t_0)$ denotes the time evolution operator from $t=t_0$ to $t=t_1$ for the single-particle problem (\ref{geq}), and $\hat{U}_0(t_1,t_0)$ is the free single-particle time evolution operator, for the Schr\"odinger problem (\ref{geq}) with potential $U(r)\to0$. Here $\hat{U}^\dagger_0(0,0)$ is of course simply 1, but we write the expression in this form in order to show that the single-particle two-time operator $\hat{S}(t_1,t_0)=\hat{U}^\dagger_0(t_1,0)\hat{U}(t_1,t_0)\hat{U}^\dagger(0,t_0)$ is simply the finite-time $S$-matrix. That is, the $S$-matrix of formal scattering theory is defined as $\hat{\mathcal{S}}=\hat{S}(\infty,\infty)$. Hence our $T(\mathbf{z},\mathbf{z}',t)$ is just the corresponding finite-time version of the $T$-matrix $\hat{\mathcal{T}}=-i(\hat{\mathcal{S}}-1)$, in the position representation, for the single-particle problem of the relative coordinate.

\subsection{Single-particle scattering}
Recognizing that equations (\ref{geq}) and (\ref{ginit}) define the propagator for a single-particle Schr\"odinger equation, we see that $g$ may be computed in the standard way, by finding a complete set of solutions to the time-\emph{independent} single-particle Schr\"odinger equation:
\begin{eqnarray}\label{gexp}
g(\mathbf{s},\mathbf{s}',t) = \int\!d^{3}k\,e^{-i\frac{\hbar}{m}k^{2}t}\psi_{\mathbf{k}}(\mathbf{s})\psi_{\mathbf{k}}^{*}(\mathbf{s}')
\end{eqnarray}
where the $\psi_{\mathbf{k}}(\mathbf{s})$ are a complete set of orthonormal energy eigenfunctions satisfying
\begin{eqnarray}\label{TISE}
	\frac{\hbar^{2}k^{2}}{m} \psi_{\mathbf{k}}(\mathbf{s}) = -\frac{\hbar^{2}}{m}\nabla_{s}^{2}\psi_{\mathbf{k}}(\mathbf{s}) + U(|\mathbf{s}|)\psi_{\mathbf{k}}(\mathbf{s}) \;.
\end{eqnarray}
The time-independent Schr\"odinger equation (\ref{TISE}) is of course a textbook problem. We assume as part of the definition of a weakly interacting gas that the range of the inter-particle potential $U$ is negligibly short compared to all other scales in this problem. For all but the negligibly small set of relative $\mathbf{s}$ at which the two particles are actually close enough to interact, the eigenfunctions $\psi_{\mathbf{k}}(\mathbf{s})$ can be written as
\begin{eqnarray}\label{partialwaves}
\psi_{\mathbf{k}}(\mathbf{s})&=& \frac{1}{(2\pi)^{3/2}}\Bigl(e^{i\mathbf{k}\cdot\mathbf{s}} + i \sum_{l=0}^{\infty}(2l+1)i^{l}P_{l}(\hat{\mathbf{k}}\cdot\hat{\mathbf{s}})\, k\,f_{l}(k) \,h_{l}(k s)\Bigr)\;,\nonumber
\end{eqnarray}
where $\hat{\mathbf{k}}$ and $\hat{\mathbf{s}}$ are the respective unit vectors in the $\mathbf{k}$ and $\mathbf{s}$ directions, $P_{l}(x)$ are the Legendre polynomials, $h_{l}=j_{l}+i n_{l}$ are the spherical Hankel functions that are composed of spherical Bessel and Neumann functions $j_{l}$ and $n_{l}$, and $f_{l}(k)$ are the complex partial wave scattering amplitudes which result from the short-ranged potential $U$. The Optical Theorem says that because the $f_{l}(k)$ do arise from a Hermitian Hamiltonian containing such a $U$, they satisfy the identity
\begin{eqnarray}\label{OT}
\mathrm{Im}\left(f_{l}(k)\right) \equiv k\, |f_{l}(k)|^{2}\;.
\end{eqnarray}
Asymptotically at large $k|\mathbf{s}|$ the Hankel function becomes a purely outgoing radial wave:
\begin{eqnarray}\label{asymptotichl}
\lim_{ks\to\infty}h_{l}(ks)=(-i)^{l+1}\frac{e^{iks}}{ks}\;,
\end{eqnarray}
and so we can recognize that at large radius $|\mathbf{s}|=s$ the scattering eigenfunctions may be written
\begin{eqnarray}\label{partialwavesAS}
\lim_{k s \to\infty}\psi_{\mathbf{k}}(\mathbf{s})&=&
\frac{1}{(2\pi)^{3/2}}\left(e^{i\mathbf{k}\cdot\mathbf{s}}+f(k,\hat{\mathbf{k}}\cdot\hat{\mathbf{s}})\frac{e^{iks}}{s}\right)\nonumber\\
f(k,\cos{\theta}) &=& \sum_{l=0}^{\infty}(2l+1)f_{l}(k)P_{l}(\cos{\theta})\;,
\end{eqnarray}
The complex scattering amplitude $f(k,\cos{\theta})$ then defines the real differential cross section $d\sigma/d\theta = |f|^{2}$ at wave number $k$.

We note for future use that although the argument $k=|\mathbf{k}| \geq 0$ in $f_{l}(k)$ is by definition non-negative, the Schr\"odinger equation (\ref{TISE}) implies a particular extension of $f_{l}(k)$ to negative $k$ arguments. We first write $\psi_{\mathbf{k}}$ from (\ref{partialwaves}) in the form
\begin{eqnarray}\label{partialwaves2}
\psi_{k,\hat{\mathbf{k}}}(\mathbf{s})&=& \frac{1}{(2\pi)^{3/2}}\Bigl(e^{ik\,\hat{\mathbf{k}}\cdot\mathbf{s}}+i\sum_{l=0}^{\infty}(2l+1)i^{l}P_{l}(\hat{\mathbf{k}}\cdot\hat{\mathbf{s}})\, k\, f_{l}(k)\,h_{l}(ks)\Bigr)\nonumber\;.
\end{eqnarray}
We then observe that since $(-k)^{2}=k^{2}$, $\psi_{-k,\hat{\mathbf{k}}}$ must also be an eigensolution to (\ref{TISE}) with energy $\hbar^{2}k^{2}/m$. Since $h_{l}(-ks)\equiv (-1)^{l}h_{l}^{*}(ks)$ is a property of spherical Hankel functions, then, we find
\begin{eqnarray}\label{partialwaves2}
\psi_{-k,\hat{\mathbf{k}}}(\mathbf{s})&=& \frac{1}{(2\pi)^{3/2}}\Bigl(e^{-ik\,\hat{\mathbf{k}}\cdot\mathbf{s}}-i\sum_{l=0}^{\infty}(2l+1)(-i)^{l}P_{l}(\hat{\mathbf{k}}\cdot\hat{\mathbf{s}})\, k\,f_{l}(-k)\,h^{*}_{l}(ks)\Bigr)\nonumber\;.
\end{eqnarray}
as another solution to (\ref{TISE}). Since (\ref{TISE}) contains no explicit factors of $i$, however, we must also recognize the solution
\begin{eqnarray}\label{partialwaves2}
\psi_{k,\hat{\mathbf{k}}}^{*}(\mathbf{s})&=& \frac{1}{(2\pi)^{3/2}}\Bigl(e^{-ik\,\hat{\mathbf{k}}\cdot\mathbf{s}}-i\sum_{l=0}^{\infty}(2l+1)(-i)^{l}P_{l}(\hat{\mathbf{k}}\cdot\hat{\mathbf{s}})\,k\,f_{l}(k)^{*}\,h^{*}_{l}(ks)\Bigr)\nonumber\;.
\end{eqnarray}
If these two solutions $\psi_{-k,\hat{\mathbf{k}}}$ and $\psi_{k,\hat{\mathbf{k}}}^{*}$ are not identical, then their difference $\psi_{-k,\hat{\mathbf{k}}}-\psi_{k,\hat{\mathbf{k}}}^{*}$ must be yet another solution; and yet it contains only incoming radial waves, with neither outgoing radial waves nor any plane wave component. Wave packets constructed out of such purely incoming radial waves would simply disappear at the origin, and this would violate unitarity. It must therefore be that $\psi_{-k,\hat{\mathbf{k}}}=\psi_{k,\hat{\mathbf{k}}}^{*}$, and therefore that
\begin{eqnarray}\label{fmink}
f_{l}(-k)\equiv f_{l}^{*}(k)\;.
\end{eqnarray}
It is easy to confirm this relation for the $f_{l}(k)$ from specific interaction potentials $U$. For the Fermi-Huang pseudopotential with s-wave scattering length $a$, for example, one has
\begin{eqnarray}\label{fminpp}
f_{l}(-k)=\frac{a\,\delta_{l0}}{1-iak}\equiv f_{l}^{*}(k)\;.
\end{eqnarray}

\subsection{The finite-time T-matrix}
Using the identities
\begin{eqnarray}\label{planewaveexp}
e^{i\mathbf{k}\cdot\mathbf{s}}&\equiv&\sum_{l=0}^{\infty}(2l+1)i^{l}j_{l}(ks)P_{l}(\hat{\mathbf{k}}\cdot\hat{\mathbf{s}})
\end{eqnarray}
and 
\begin{eqnarray}\label{Legendreintegral}
\oint\!d^{2}\Omega_{k}\,P_{l}(\hat{\mathbf{k}}\cdot\hat{\mathbf{s}})P_{l'}(\hat{\mathbf{k}}\cdot\hat{\mathbf{s}}')&\equiv&\frac{4\pi\delta_{ll'}}{2l+1}P_{l}(\hat{\mathbf{s}}\cdot\hat{\mathbf{s}}')
\end{eqnarray}
we can compute $g(\mathbf{s},\mathbf{s}',t)$ more explicitly by performing the angular part $\oint d^{2}\Omega_{k}$ of the $\int d^{3}k$ integration in (\ref{gexp}). We obtain
\begin{eqnarray}\label{gcomp}
g(\mathbf{s},\mathbf{s}',t)&=& \left(\frac{m}{4\pi i \hbar t}\right)^{\frac{3}{2}}e^{\frac{i m}{4\hbar t}|\mathbf{s}-\mathbf{s}'|^{2}} 
			+ \frac{i}{(2\pi)^{2}}\sum_{l=0}^{\infty}(2l+1)P_{l}(\hat{\mathbf{s}}\cdot\hat{\mathbf{s}}')\int_{0}^{\infty}\!k^{2}dk\,e^{-i\frac{\hbar}{m}k^{2}t}F_{l}(ks,ks')\nonumber\\
			F_{l}(ks,ks')
			& \equiv& f_{l}(k)h_{l}(ks)h_{l}(ks')+ f_{l}(-k)h_{l}(-ks)h_{l}(-ks')\;,
\end{eqnarray}
where we obtain the second expression by using the Optical Theorem (\ref{OT}) and the definition $h_{l}=j_{l}+in_{l}$ of the Hankel functions, and in the last line we have used the definition (\ref{fmink}) of $f_{l}(-k)$. Note that the integrand $F_{l}$ really does include the product $h_{l}(ks)h_{l}(ks')$, and \textit{not} $h_{l}(ks)h_{l}^{*}(ks')$; this reflects the fact that $g(\mathbf{s},\mathbf{s}',t)$ is unitary rather than Hermitian. $F_{l}$ is also correctly expressed as given in its final form with $f_{l}(k)$ but no $|f_{l}(k)|^{2}$, even though it was obtained from a $\psi_{\mathbf{k}}^{*}\psi_{\mathbf{k}}$ integrand that did include such an $|f_{l}(k)|^{2}$ cross term; the exact cancelation of the $|f_{l}(k)|^{2}$ term was due to the Optical Theorem, which is itself again due to the unitarity of $g(\mathbf{s},\mathbf{s}',t)$.

Collecting the results in (\ref{gcomp}) we obtain a surprisingly compact expression for the single-particle propagator, which is nonetheless exactly valid everywhere outside the short range of the interaction potential $U$:
\begin{eqnarray}\label{gcomp2}
g(\mathbf{s},\mathbf{s}',t)&=& \left(\frac{m}{4\pi i \hbar t}\right)^{\frac{3}{2}}e^{\frac{i m}{4\hbar t}|\mathbf{s}-\mathbf{s}'|^{2}} 
			+ \frac{i}{(2\pi)^{2}}\sum_{l=0}^{\infty}(2l+1)P_{l}(\hat{\mathbf{s}}\cdot\hat{\mathbf{s}}')\int_{-\infty}^{\infty}\!k^{2}dk\,e^{-i\frac{\hbar}{m}k^{2}t} f_{l}(k)h_{l}(ks)h_{l}(ks')\;.
\end{eqnarray}
Using the asymptotic behavior of the Hankel functions (\ref{asymptotichl}), we find that this has the even simpler asymptotic limit
\begin{eqnarray}\label{gAS}
\lim_{s,s'\to\infty}g(\mathbf{s},\mathbf{s}',t)&=& \left(\frac{m}{4\pi i \hbar t}\right)^{\frac{3}{2}}e^{\frac{i m}{4\hbar t}|\mathbf{s}-\mathbf{s}'|^{2}} 
			- \frac{i}{(2\pi)^{2}}\int_{-\infty}^{\infty}\!dk\,k\,f(k,-\hat{\mathbf{s}}\cdot\hat{\mathbf{s}}')e^{-i\frac{\hbar}{m}k^{2}t} \frac{e^{ik(s+s')}}{s\,s'}\;,
\end{eqnarray}
using the parity of the Legendre polynomials $P_{l}(-x)=(-1)^{l}P_{l}(x)$. This confirms that the $f_{l}(k)$ in our exact propagator (\ref{gcomp2}) do provide the standard scattering effect of $f(k,\cos\theta)$ in the limit of infinite times and long range.

Inserting (\ref{gcomp2}) in the definition of $T(\mathbf{s},\mathbf{s}',t)$ in the second line of (\ref{RelationL2toT}), we obtain 
\begin{eqnarray}\label{Tresult}
T(\mathbf{s},\mathbf{s}',t)&=&\left(\frac{im}{4\pi\hbar t}\right)^{\frac{3}{2}}\int\!d^3\,s''\,e^{-i\frac{m}{4\hbar t}|\mathbf{s}-\mathbf{s}''|^2}\sum_{l=0}^{\infty}\frac{(2l+1)P_{l}(\hat{\mathbf{s}}''\cdot\hat{\mathbf{s}}')}{(2\pi)^{2}}\int_{-\infty}^{\infty}\!k^{2}dk\,e^{-i\frac{\hbar}{m}k^{2}t}f_{l}(k)h_{l}(ks'')h_{l}(ks')
\end{eqnarray}
exactly, everywhere outside the range of the interaction potential $U$.

NOT CLEAR THAT WE NEED THIS LAST PARAGRAPH OR THE WHOLE NEXT APPENDIX.
Since colliding particles interact briefly and then separate, it is only the behavior of $g(\mathbf{r}-\mathbf{r}',\mathbf{q}_{1}-\mathbf{q}_{2},t)$ for large magnitudes of its spatial arguments that can represent the steadily accumulating long-term effects of collisions. We are therefore only interested in the asymptotic form of $\psi_{\mathbf{k}}(\mathbf{s})$ at large $|\mathbf{s}|$. Indeed, the neglect of everything except asymptotic long-distance behavior in $\psi_{\mathbf{k}}(\mathbf{s})$ is not really an approximation, but rather an exact statement of the specific question that we are asking about gas dynamics, concerning the long-term effects, without transients.

\subsection{Derivation of the integral representation of $g(\mathbf{s},\mathbf{s}',t)$ in Equation (\ref{gAS})}\label{AppendixPartialWaves}

Inserting (\ref{partialwavesAS}) into (\ref{gexp}) for $k|\mathbf{s}|$ and $k|\mathbf{s}'|$ both large, we find four different combinations of the incident plane waves and scattered radial waves:
\begin{eqnarray}\label{gexp2}
\lim_{s,s'\to\infty}g(\mathbf{s},\mathbf{s}',t) &=& \frac{1}{(2\pi)^{3}}\!\!\int\!d^{3}k\,e^{-\frac{i\hbar k^{2}t}{m}}\left[e^{i\mathbf{k}\cdot(\mathbf{s}-\mathbf{s}')} + e^{i\mathbf{k}\cdot\mathbf{s}}\frac{e^{-iks'}}{s'}f^{*}(k,\mathbf{\mathbf{\hat{k}}}\cdot\mathbf{\hat{s}}') + e^{-i\mathbf{k}\cdot\mathbf{s}'}\frac{e^{iks}}{s}f(k,\mathbf{\hat{k}}\cdot\mathbf{\hat{s}})  \right.  \nonumber\\
& & \,\left.\qquad\qquad\qquad\qquad\qquad\qquad\qquad\qquad +  \frac{e^{ik(s-s')}}{s's} f^{*}(k,\mathbf{\hat{k}}\cdot\mathbf{\hat{s}}')f(k,\mathbf{\hat{k}}\cdot\mathbf{\hat{s}})\right]
\end{eqnarray}
where $\mathbf{\mathbf{\hat{k}}}$, $\mathbf{\hat{s}}$, and $\mathbf{\hat{s}}'$ are not operators, but unit vectors in the directions of $\mathbf{k}$, $\mathbf{s}$, and $\mathbf{s}'$, respectively. We then apply the expansion of a plane wave in spherical harmonics and spherical Bessel functions $j_{l}(kr)$ \cite{Tayl72},
 \begin{eqnarray}\label{plane2Y}
 e^{i\mathbf{k}\cdot\mathbf{r}}&=&\sum_{l=0}^{\infty}(2l+1)i^{l}j_{k}(kr)\,P_{l}(\mathbf{\hat{k}}\cdot\mathbf{\hat{r}})\nonumber\underset{kr\gg1}{\longrightarrow}\sum_{l=0}^{\infty}(2l+1)\frac{e^{ikr}-(-1)^{l}e^{-ikr}}{2ikr}P_{l}(\mathbf{\hat{k}}\cdot\mathbf{\hat{r}})\;,
 \end{eqnarray}
to the second and third terms in the integrand of (\ref{gexp2}), to obtain
\begin{eqnarray}\label{gexp3}
\lim_{s,s'\to\infty}g(\mathbf{s},\mathbf{s}',t) &=& \left(\frac{m}{4\pi i \hbar t}\right)^{\frac{3}{2}}e^{\frac{i m}{4\hbar t}|\mathbf{s}'-\mathbf{s}|^{2}}\nonumber\\
&& + \int_{0}^{\infty}\!\frac{dk}{(2\pi)^{3}}\,\frac{e^{-\frac{i\hbar k^{2}t}{m}}}{s's}\left[e^{ik(s-s')}\,\alpha_{k}(\mathbf{\hat{s}},\mathbf{\hat{s}}')-i\left(e^{ik(s+s')}\beta_{k}(\mathbf{\hat{s}},\mathbf{\hat{s}}')-e^{-ik(s+s')}\beta^{*}_{k}(\mathbf{\hat{s}}',\mathbf{\hat{s}})\right)\right]
\end{eqnarray}
where the factors $\alpha_{k}$ and $\beta_{k}$ contain the integrals over $\mathbf{k}$-space angular coordinates 
\begin{eqnarray}\label{gammak}
\alpha_{k}(\mathbf{\hat{s}},\mathbf{\hat{s}}') &=& \oint\!d^{2}\Omega_{k}\,\left[k^2\,f^{*}(k,\mathbf{\hat{k}}\cdot\mathbf{\hat{s}}')f(k,\mathbf{\hat{k}}\cdot\mathbf{\hat{s}}) -i\,\frac{k}{2}\sum_{l=0}^{\infty}(2l+1)[P_{l}(\mathbf{\hat{k}}\cdot\mathbf{\hat{s}})f^{*}(k,\mathbf{\hat{k}}\cdot\mathbf{\hat{s}}')-P_{l}(\mathbf{\hat{k}}\cdot\mathbf{\hat{s}}')f(k,\mathbf{\hat{k}}\cdot\mathbf{\hat{s}})]\right]\nonumber\\
\beta_{k}(\mathbf{\hat{s}},\mathbf{\hat{s}}')&=& \frac{k}{2}\sum_{l=0}^{\infty}(-1)^{l}(2l+1)\oint\!d^{2}\Omega_{k}\,f(k,\mathbf{\hat{k}}\cdot\mathbf{\hat{s}})P_{l}(\mathbf{\hat{k}}\cdot\mathbf{\hat{s}}')\;.
\end{eqnarray}

Inserting the expansion (\ref{partialwavesAS}) of $f$ in Legendre polynomials we can show that all terms in $\alpha_{k}$ and $\beta_{k}$ involve the same kind of angular integral:
\begin{eqnarray}\label{alphabeta}
\alpha_{k}(\mathbf{\hat{s}},\mathbf{\hat{s}}') &=& k\,\sum_{l,l'=0}^{\infty}(2l+1)(2l'+1)\Big(k\,f_{l}(k)f^{*}_{l'}(k)-\frac{i}{2}[f^{*}_{l'}(k)-f_{l}(k)]\Big)\oint\!d^{2}\Omega_{k}\,P_{l}(\mathbf{\hat{k}}\cdot\mathbf{\hat{s}})P_{l'}(\mathbf{\hat{k}}\cdot\mathbf{\hat{s}}')\nonumber\\
\beta_{k}(\mathbf{\hat{s}},\mathbf{\hat{s}}')&=&\frac{k}{2}\sum_{l,l'=0}^{\infty}(-1)^{l'}(2l+1)(2l'+1)f_{l}(k)\oint\!d^{2}\Omega_{k}\,P_{l}(\mathbf{\hat{k}}\cdot\mathbf{\hat{s}})P_{l'}(\mathbf{\hat{k}}\cdot\mathbf{\hat{s}}')\;.
\end{eqnarray}

We can evaluate these angular integrals using the so-called  \emph{addition theorem for spherical harmonics}\cite{addYlm}
\begin{eqnarray}\label{YlmAdd}
P_{l}(\mathbf{\hat{r}}\cdot\mathbf{\hat{r}}') = \frac{4\pi}{2l+1}\sum_{m=-l}^{l}Y^{*}_{lm}(\theta_{r'},\phi_{r'})Y_{lm}(\theta_{r},\phi_{r})\;,
\end{eqnarray}
where for any vector $\mathbf{v}$ we define polar coordinates $\theta_{v},\phi_{v}$ such that the unit vector in the direction of $\mathbf{v}$ is $\hat{v}=(\sin\theta_{v}\cos\phi_{v},\sin\theta_{v}\sin\phi_{v},\cos\theta_{v})$.
Then, using the fact that the $Y_{lm}$ are orthonormal on the unit sphere
\begin{eqnarray}\label{Ylmortho}
\oint\!d^{2}\Omega\,Y^{*}_{lm}(\theta,\phi)\,Y_{l'm'}(\theta,\phi) = \delta_{ll'}\delta_{mm'}\;,
\end{eqnarray}
we can derive the identity
\begin{eqnarray}\label{ID2}
\oint\!d^{2}\Omega_{k}\,P_{l'}(\mathbf{\hat{k}}\cdot\mathbf{\hat{s}})P_{l}(\mathbf{\hat{k}}\cdot\mathbf{\hat{s}}') &=& \frac{(4\pi)^{2}}{(2l+1)(2l'+1)}\sum_{m=-l}^{l}\sum_{m'=-l'}^{l'}Y_{lm}^{*}(\theta_{s'},\phi_{s'})Y_{l'm'}(\theta_{s},\phi_{s})\oint\!d^{2}\Omega_{k}\,Y_{lm}(\theta_{k},\phi_{k})Y_{l'm'}^{*}(\theta_{k},\phi_{k})\nonumber\\
&=&\frac{(4\pi)^{2}}{(2l+1)(2l'+1)}\sum_{m=-l}^{l}\sum_{m'=-l'}^{l'}Y_{lm}^{*}(\theta_{s'},\phi_{s'})Y_{l'm'}(\theta_{s},\phi_{s})\delta_{ll'}\delta_{mm'}\nonumber\\
&=&\frac{(4\pi)^{2}\delta_{ll'}}{(2l+1)^{2}}\sum_{m=-l}^{l}Y_{lm}^{*}(\theta_{s'},\phi_{s'})Y_{lm}(\theta_{s},\phi_{s})\equiv \frac{4\pi\delta_{ll'}}{(2l+1)}P_{l}(\mathbf{\hat{s}}\cdot\mathbf{\hat{s}}')\;,
\end{eqnarray}
where for the final equality we again apply the addition theorem (\ref{YlmAdd}).

Applying the identity (\ref{ID2}) to (\ref{alphabeta}) then reveals
\begin{eqnarray}
\alpha_{k}(\mathbf{\hat{s}},\mathbf{\hat{s}}') &\equiv& -k\left[4\pi\mathrm{Im}[f(k,\mathbf{\hat{s}}\cdot\mathbf{\hat{s}}')]- k\,\oint\!d^{2}\Omega_{k}\,f^{*}(k,\mathbf{\hat{k}}\cdot\mathbf{\hat{s}}')f(k,\mathbf{\hat{k}}\cdot\mathbf{\hat{s}})\right]\label{alphaA}\\
&\equiv & 4\pi\sum_{l=0}^{\infty}(2l+1)\Big(k\,|f_{l}(k)|^{2}-\mathrm{Im}[f_{l}(k)]\Big)\label{alphaB}\\
\beta_{k}(\mathbf{\hat{s}},\mathbf{\hat{s}}')&=& 2\pi k\sum_{l=0}^{\infty}(-1)^{l}(2l+1)f_{l}(k)P_{l}(\mathbf{\hat{s}}\cdot\mathbf{\hat{s}}') \equiv 2\pi \,k\, f(k,-\mathbf{\hat{s}}\cdot\mathbf{\hat{s}}')\;,\label{betaB}
\end{eqnarray}
where in the last equality we have used the parity property of Legendre polynomials, $P_{l}(-x)=(-1)^{l}P_{l}(x)$.

We now use the fact that the $f_{l}(k)$ are not completely arbitrary complex numbers, but are constrained by the unitarity of wave mechanical evolution to have the form $f_{l}(k) = e^{i\delta_{l}(k)}\sin\delta_{l}(k)$ for some set of real partial wave phase shifts $\delta_{l}(k)$ \cite{CTPS}. This implies that $k|f_{l}(k)|^{2}\equiv \mathrm{Im}[f_{l}(k)]$ for all $k$ and $l$, so that $\alpha_{k}$ vanishes identically according to (\ref{alphaB}). Combining (\ref{betaB}) with $\alpha_{k}=0$ brings (\ref{gexp3}) into the simple form 
\begin{eqnarray}
\lim_{s,s'\to\infty}g(\mathbf{s},\mathbf{s}',t) &=& \left(\frac{m}{4\pi i \hbar t}\right)^{\frac{3}{2}}e^{\frac{i m}{4\hbar t}|\mathbf{s}'-\mathbf{s}|^{2}}
-\frac{i}{(2\pi)^{2}s's} \int_{0}^{\infty}\!dk\, k\,e^{-\frac{i\hbar k^{2}t}{m}}\left[f(k,-\mathbf{\hat{s}}\cdot\mathbf{\hat{s}}')e^{ik(s+s')}-f^{*}(k,-\mathbf{\hat{s}}\cdot\mathbf{\hat{s}}')e^{-ik(s+s')}\right]\;.\nonumber\\
\end{eqnarray}
With the definition $f(-k,\cos\theta) = f^{*}(k,\cos\theta)$, as given in our main text, this is equivalent to Eqn.~(\ref{gpre}). 

When $\alpha_{k}$ is expressed in the form (\ref{alphaA}), we can see that the fact that $\alpha_{k}$ vanishes identically due to unitarity means that
\begin{eqnarray}
 \oint\!d^{2}\Omega_{k}\,f^{*}(k,\mathbf{\hat{k}}\cdot\mathbf{\hat{s}}')f(k,\mathbf{\hat{k}}\cdot\mathbf{\hat{s}})\equiv\frac{4\pi}{k}\mathrm{Im}[f(k,\mathbf{\hat{s}}\cdot\mathbf{\hat{s}}')]\;,\nonumber\\
 \end{eqnarray}
 for any two unit vectors $\mathbf{\hat{s}}$ and $\mathbf{\hat{s}}'$. The identity $\alpha_{k}(\mathbf{\hat{s}},\mathbf{\hat{s}}')\equiv 0$ is known as the \textit{generalized optical theorem} \cite{Tayl72}.

\end{widetext}

\end{appendix}


\begin{thebibliography}{XXXXXXXXXXX}
\bibitem{Nord28} L.W.~Nordheim, Proc. Roy. Soc. London, Ser. A, {\bf 119}, 689 (1928).
\bibitem{Tolm38} R.C.~Tolman, {\it The principles of statistical mechanics}, Oxford Univ. Press, London (1938).
\bibitem{JoosZeh85} E.~Joos and H.D.~Zeh, Z. Phys. B {\bf 59}(2), 223-243 (1985).
\bibitem{Greenwood} D.A.~Greenwood, Proc. Phys. Soc. {\bf 71}, 585 (1958), is forthright about the phenomenological nature of this step. Coherences in the momentum representation of the single-particle correlation function are discarded on page 3 of this 11-page paper, with an appeal to randomization of phases, and the step is admitted to be a `` \textit{Stosszahlansatz} procedure'' which must be accepted, rather than derived. 
\bibitem{Tayl72} J.R.~Taylor, {\it Scattering Theory: The quantum theory of nonrelativistic collisions}, John Wiley \& Sons, New York (1972).
\bibitem{CaldeiraLeggett83} A.O. Caldeira and A.J. Leggett, Physica A {\bf 121}, 587 (1983).
\bibitem{addYlm}G.~Arfken, \textit{Mathematical Methods for Physicists}, 3rd ed. (Orlando, FL: Academic Press, 1985), pp. 693-695.
\bibitem{Kada89} L.P.~Kadanoff and G.~Baym, {\it Quantum Statistical Mechanics: Green's function methods in equilibrium and nonequilibrium problems}, Perseus Books, Cambridge (1989).
\bibitem{GallisFleming90} M.R.~Gallis and G.N.~Fleming, Phys. Rev. A {\bf 42}, 38 (1990).
\bibitem{Born96} T.~Bornath, D.~Kremp, W.D.~Kraeft and M.~Schlanges, Phys. Rev. E {\bf 54}, 3274 (1996).
\bibitem{Gard97} C.W.~Gardiner and P.~Zoller, Phys. Rev. A {\bf 55}, 2902 (1997).
\bibitem{AnglinPazZurek97} J.R.~Anglin, J.P.~Paz, and W.H.~Zurek, Phys. Rev. A {\bf 55}, 4041 (1997).
\bibitem{ZNGr98} E.~Zaremba, A.~Griffin and T.~Nikuni, Phys. Rev. A {\bf 57}, 4695 (1998).
\bibitem{BenderOrszag99} C. M. Bender and S. A. Orszag, {\it Advanced mathematical methods for scientists and engineers I: asymptotic methods and perturbation theory}, Springer, New York (1999)
\bibitem{Kett02} J.M.~Vogels, K.~Xu, and W.~Ketterle, Phys. Rev. Lett. {\bf 89}, 020401 (2002).
\bibitem{Zurek03} W.H.~Zurek, Rev. Mod. Phys. {\bf 75}, 715 (2003).
\bibitem{Ott04} C.~Modugno, E.~de Mirandes, F.~Ferlaino, H.~Ott, G.~Roati and M.~Inguscio, Fortschr. Phys. {\bf 52}, 1173 (2004)
\bibitem{Horn06} K.~Hornberger, Phys. Rev. Lett. {\bf 97}, 060601 (2006).
\bibitem{Pethick08} C. J. Pethick and H. Smith, {\it Bose-Einstein condensation in dilute gases}, 2nd edition, Cambridge University Press (2008)
\bibitem{Gilz14} L. Gilz, L. Rico-P\'erez and J.R. Anglin, Phys. Rev. A {\bf 89}, 052131 (2014).
\bibitem{CTPS} C. Cohen-Tannoudji, B. Diu and F. Lalo\"e, \textit{Quantum Mechanics} (Wiley, 1977), p. 934.
\bibitem{Cercignani75} C. Cercignani, \textit{Theory and Application of the Boltzmann Equation}, Scottish Academic Press (1975)
\bibitem{Kim2017} Hui Dong, Da-wei Wang, M.B. Kim, arXiv 1706.02636 (2017).
\end{thebibliography}
\end{document}